\documentclass[a4paper,twocolumn,11pt]{quantumarticle}
\pdfoutput=1
\usepackage[utf8]{inputenc}
\usepackage[english]{babel}
\usepackage[T1]{fontenc}
\usepackage{amsmath}
\usepackage{hyperref}
\usepackage{physics}
\usepackage{tikz}
\usepackage{lipsum}
\usepackage{booktabs}
\usepackage{caption}
\usepackage{subcaption}
\usepackage{graphicx}
\usepackage{verbatim}

\begin{document}

\title{Mapping Guaranteed Positive Secret Key Rates for Continuous Variable Quantum Key Distribution}

\author{Mikhael Sayat}
\affiliation{Department of Physics, Faculty of Science, University of Auckland, Auckland 1010, New Zealand}
\affiliation{Department of Quantum Science and Technology, Research School of Physics, Australian National University, Canberra, ACT 2601, Australia}
\author{Oliver Thearle}
\affiliation{Department of Quantum Science and Technology, Research School of Physics, Australian National University, Canberra, ACT 2601, Australia}
\affiliation{Sensors and Effectors Division, Defence Science and Technology Group, Adelaide 5111, Australia}
\author{Biveen Shajilal}
\affiliation{A*STAR Quantum Innovation Centre (Q.InC), Institute of Materials Research and Engineering (IMRE), Agency for Science, Technology and Research (A*STAR), 2 Fusionopolis Way, 08-03 Innovis 138634, Singapore}
\author{Sebastian P. Kish}
\affiliation{Data61, Commonwealth Scientific and Industrial Research Organisation, Sydney, NSW 2015, Australia}
\author{Ping Koy Lam}
\affiliation{A*STAR Quantum Innovation Centre (Q.InC), Institute of Materials Research and Engineering (IMRE), Agency for Science, Technology and Research (A*STAR), 2 Fusionopolis Way, 08-03 Innovis 138634, Singapore}
\affiliation{Department of Quantum Science and Technology, Research School of Physics, Australian National University, Canberra, ACT 2601, Australia}
\author{Nicholas Rattenbury}
\affiliation{Department of Physics, Faculty of Science, University of Auckland, Auckland 1010, New Zealand}
\author{John Cater}
\affiliation{Aerospace Research Institute, University of Canterbury, Christchurch 8140, New Zealand}

\maketitle
\onecolumn
\begin{abstract}
Continuous variable quantum key distribution (CVQKD) is the sharing of secret keys between different parties using the continuous amplitude and phase quadratures of light. There are many protocols in which different modulation schemes are used to implement CVQKD. However, there has been no tool for comparison between different CVQKD protocols to determine the optimal protocol for varying channels while simultaneously taking into account the effects of different parameters. Here, a comparison tool has been developed to map regions of positive secret key rate (SKR), given a channel's transmittance and excess noise, where a user's modulation can be adjusted to guarantee a positive SKR in an arbitrary environment. The method has been developed for discrete modulated CVQKD (DM-CVQKD) protocols but can be extended to other current and future protocols and security proofs.
\vspace{-1mm}
\end{abstract}

\section{Introduction}
Quantum key distribution (QKD) is the sharing of keys between different parties, Alice and Bob, where the presence of an eavesdropper, Eve, can be inferred as a consequence of fundamental quantum mechanics \cite{Bennett1984, Scarani2009}. Continuous variable QKD (CVQKD) uses the phase and amplitude of a laser to encode information that is used for QKD \cite{Braunstein2005, Weedbrook2012}.

The common CVQKD protocol, GG02 \cite{Grosshans2002} modulates the phase and amplitude of a laser using a Gaussian distribution to produce light similar to half of a two mode squeeze vacuum. An alternative is to modulate the phase and amplitude by a discretised distribution. There have been a number of different protocols using discrete modulation published \cite{Leverrier2009, Becir2010, Zhang2012, Denys2021, Almeida2021}. The modulated light is sent through a channel to Bob who, with the help of Alice, can derive a shared secret from their shared light. The channel is typically modelled as a Gaussian loss channel. The main advantage of discrete modulated CVQKD (DM-CVQKD) is that it can tolerate more noise, and therefore has better performance in low signal-to-noise (SNR) regimes, as the coherent states are more distinguishable compared to Gaussian modulated CVQKD (GM-CVQKD).

 For a given channel with fixed transmittance and excess noise, a secret key rate (SKR) can be calculated. In the case of a real channel, channel parameters (transmittance and excess noise) vary between a CVQKD transmitter and receiver. The calculation of a SKR is generally inefficiently performed by brute force by varying one channel parameter (transmittance or excess noise) \cite{Becir2010, Zhang2012, Laudenbach2018, Kish2020}. 
 
 In this work, an efficient tool for comparing different CVQKD protocols is developed for different channels by calculating the boundary between positive and negative SKRs to identify the CVQKD parameters required to guarantee a positive SKR. The tool provides a map with regions of positive and negative SKRs for channel parameters (transmittance and excess noise) and the required modulation amplitude ($\alpha$) of different CVQKD protocols to guarantee a positive SKR. Although the tool involves the calculation of SKRs to establish the boundary, it is an assessment of how different CVQKD parameters simultaneously affect the capability to produce positive SKRs and performance of different protocols without explicitly showing the SKR.

Amongst the many DM-CVQKD protocols (original and extensions for optimisation) the following protocols are considered \cite{Denys2021, Almeida2021}: 
\begin{itemize}
    \item $M$ - Phase Shift Keying ($M$-PSK),
    \item $M$ - Quadrature Amplitude Modulation ($M$-QAM), and
    \item $M$ - Amplitude and Phase Shift Keying ($M$-APSK).
\end{itemize}
Here $M$ denotes the number of discrete levels used in the modulation protocol.

As a universal finite size limit SKR security proof for DM-CVQKD is still an active field of research \cite{Matsuura2021, Yamano2022, Kanitschar2023}, the analysis has been restricted to the asymptotic limit. In addition, the focus of this work is the tool and not the underlying protocols and security proofs.

\section{DM-CVQKD Protocols}
In CVQKD, information is encoded in the two quadratures of light by modulating coherent states in the amplitude ($x$) and phase quadratures ($p$). A series of coherent states, each one represented by the notation ($\ket{\alpha} = \ket{x + ip}$), are sent from Alice to Bob where they are measured. If Eve was to intercept any of the transmitted states then their presence would appear as noise and loss to Alice and Bob due to the no cloning theorem \cite{Grosshans2002}. The distribution of these coherent states on the phase space are discrete for DM-CVQKD. The different modulation schemes are described in this section. 


The SKR is the lower bound on the key rate calculated by Alice and Bob by determining their shared information ($I_{AB}$) and subtracting the upper bound of the inferred information intercepted by Eve ($S_{BE}$). $I_{AB}$ can be written as

\begin{equation}
    \label{SKR}
    \mathrm{SKR} = \beta I_{AB} - S_{BE},
\end{equation}
where $\beta$ is the reconciliation efficiency, a method of error-correction on the transmitted and received coherent states \cite{Laudenbach2018}.

The parameters and SKR can be calculated in two ways. The first method uses a linear channel assuming (LCA) security analysis \cite{Leverrier2009, Zhang2012, Wang2022}. The second uses a semi-definite programming (SDP) method \cite{Denys2021, Wang2022, Ghorai2019, Lin2019} that requires more computational power but has a more general secure analysis theory than the LCA method. 

In developing the tool, the calculation of the SKR for the three DM-CVQKD protocols employs the analytical LCA method from Denys et al. \cite{Denys2021} which is secure against collective attacks; which are a form of attack from Eve where Eve uses ancilla states for an independent and identically distributed attack, stores the state in a quantum memory, and later performs an optimal collective measurement \cite{Laudenbach2018}. The rest of this section will describe the distribution of coherent states on the phase space for $M$-PSK, $M$-QAM, and $M$-APSK. 

\subsection{$M$-PSK}
\begin{figure}[htp]
     \centering
     \begin{subfigure}[b]{0.43\textwidth}
         \centering
         \includegraphics[width=\textwidth]{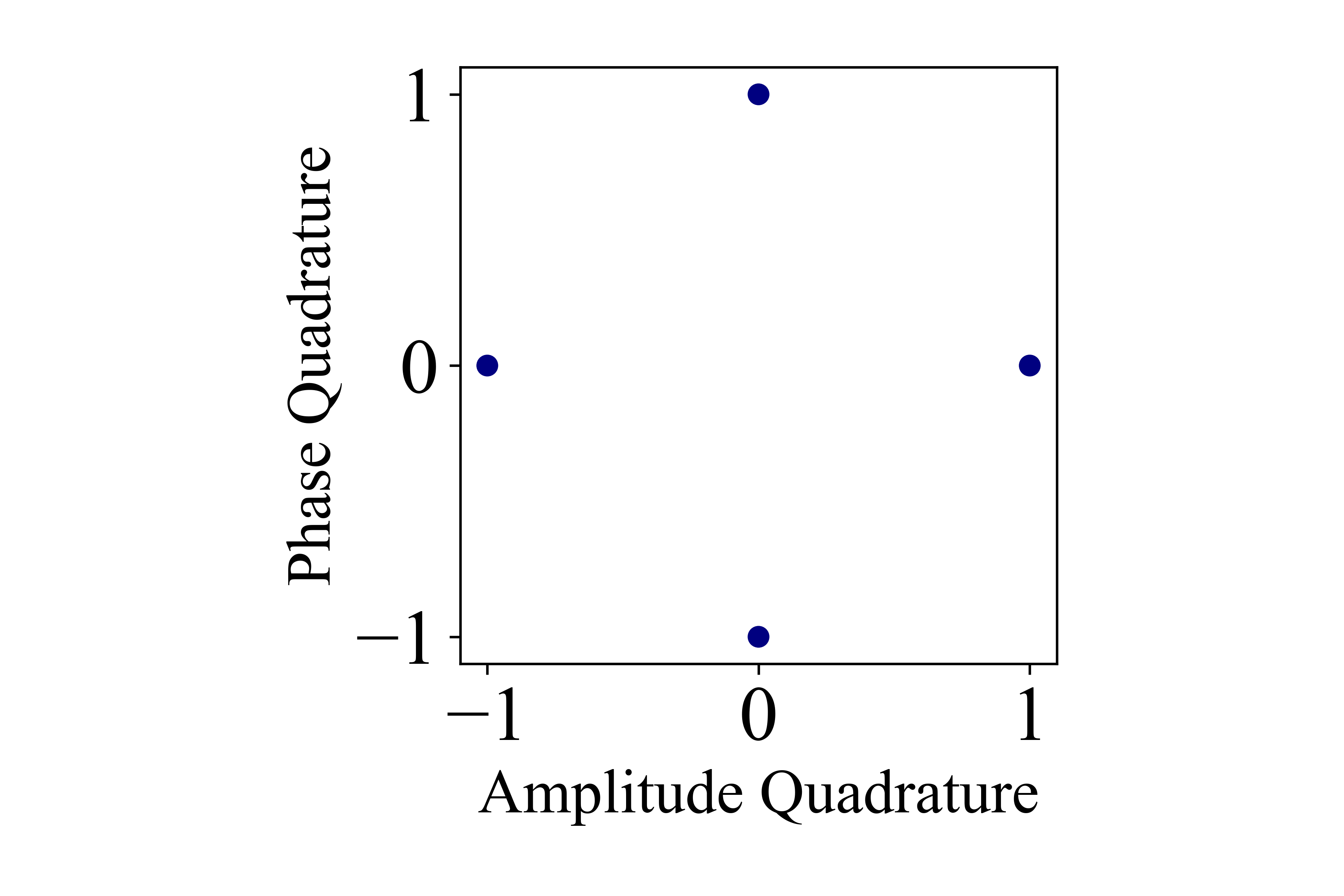}
         \caption{4-PSK}
         \label{fig:4PSK}
     \end{subfigure}
     \hspace{-3.5em}
     \begin{subfigure}[b]{0.43\textwidth}
         \centering
         \includegraphics[width=\textwidth]{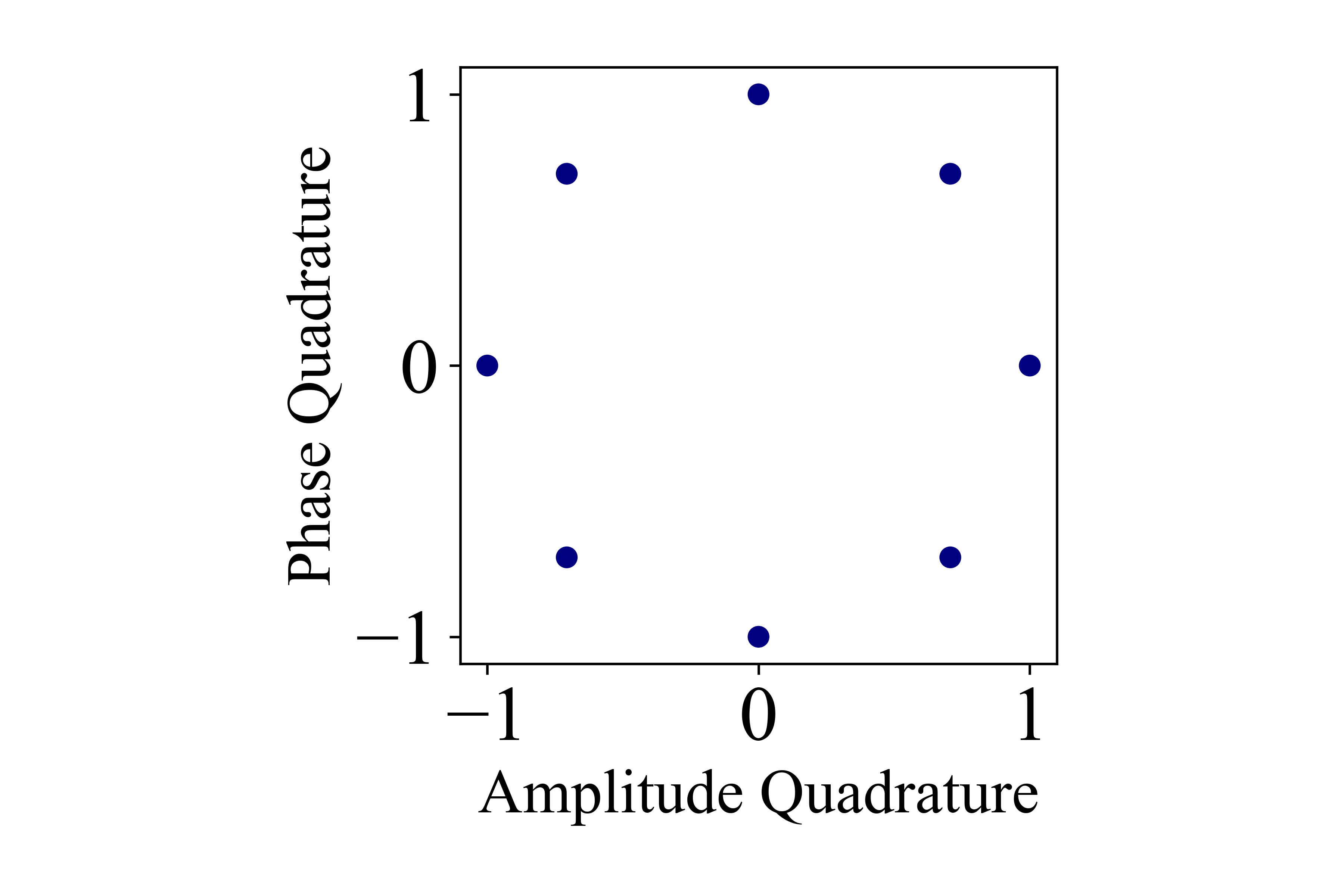}
         \caption{8-PSK}
         \label{fig:8PSK}
     \end{subfigure}
     \begin{subfigure}[b]{0.43\textwidth}
         \centering
         \includegraphics[width=\textwidth]{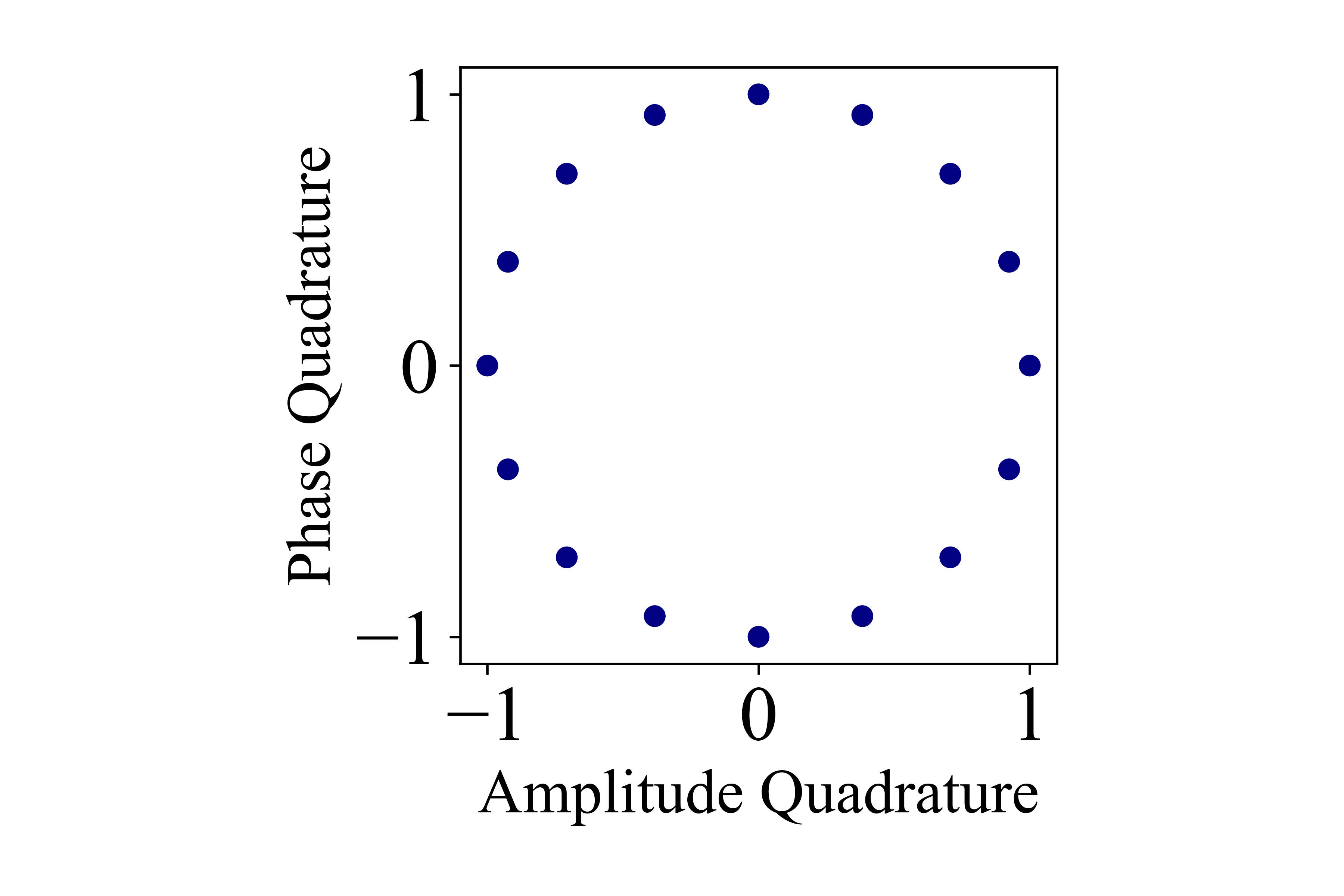}
         \caption{16-PSK}
         \label{fig:16PSK}
     \end{subfigure}
     \hspace{-3.5em}
     \begin{subfigure}[b]{0.43\textwidth}
         \centering
         \includegraphics[width=\textwidth]{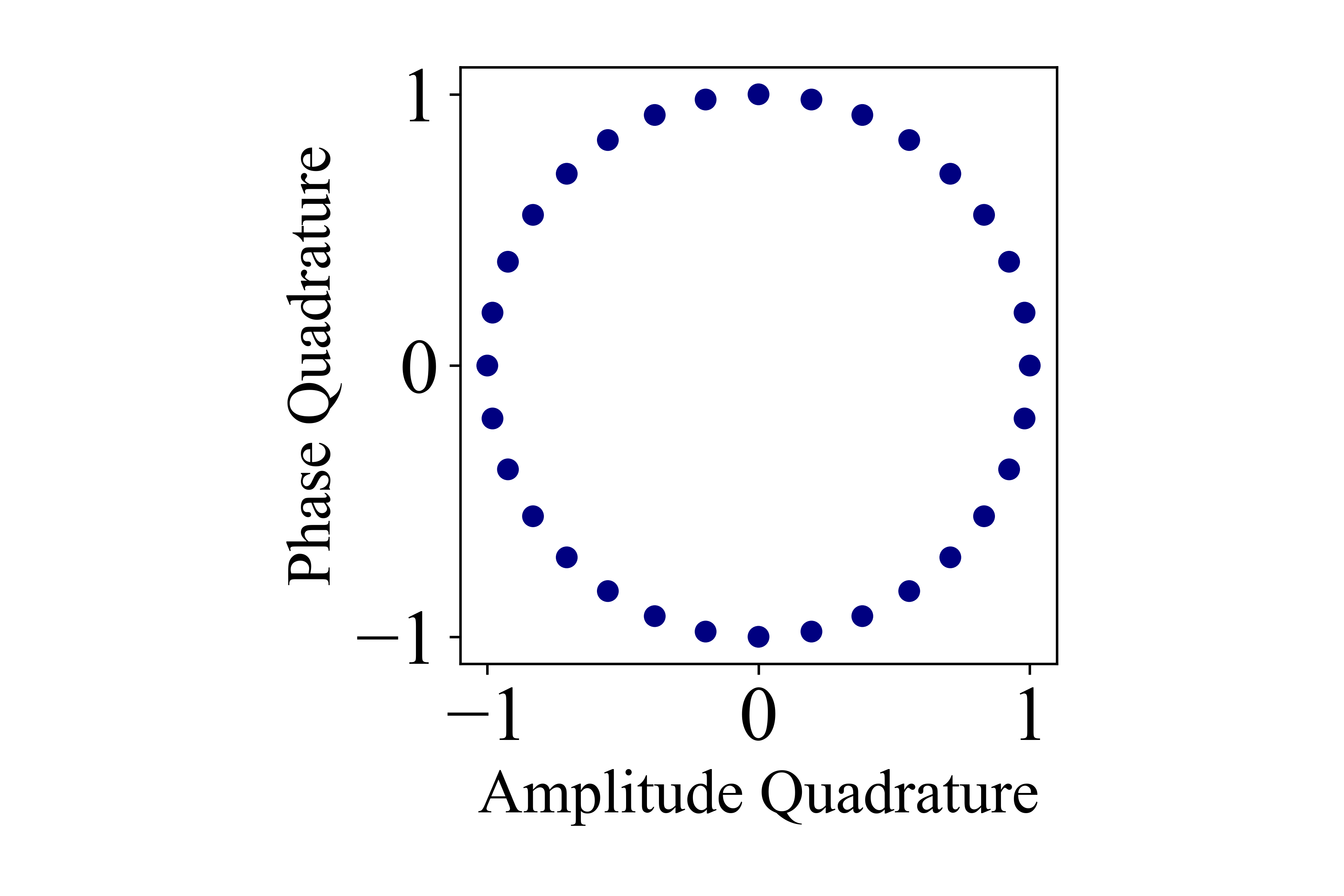}
         \caption{32-PSK}
         \label{fig:32PSK}
     \end{subfigure}
    \caption{Phase space representation of 4,8,16,32-PSK protocols, $\alpha = 1$. Each coherent state has uniform probability $\left(\frac{1}{M}\right)$. }
    \label{fig:M-PSK}
\end{figure}

$M$-PSK is a distribution of $M$ modulated coherent states on the phase space with a constant value of $\alpha$ and a uniform probability $\left( \frac{1}{M} \right)$ \cite{Zhang2012, Djordjevic2019}. A modulated coherent state takes the form

\begin{equation}
    \ket{\alpha_k} = \alpha e^{\frac{i2\pi k}{M}},
\end{equation}
and can be depicted on the phase space as shown in Figure \ref{fig:M-PSK}.

\subsection{$M$-QAM}
$M$-QAM is a distribution of $M$ modulated coherent states on the phase space following a grid-like pattern where each coherent state is assigned a different probability \cite{Denys2021}. 

\begin{figure}[htp!]
     \centering
     \begin{subfigure}[b]{0.43\textwidth}
         \centering
         \includegraphics[width=\textwidth]{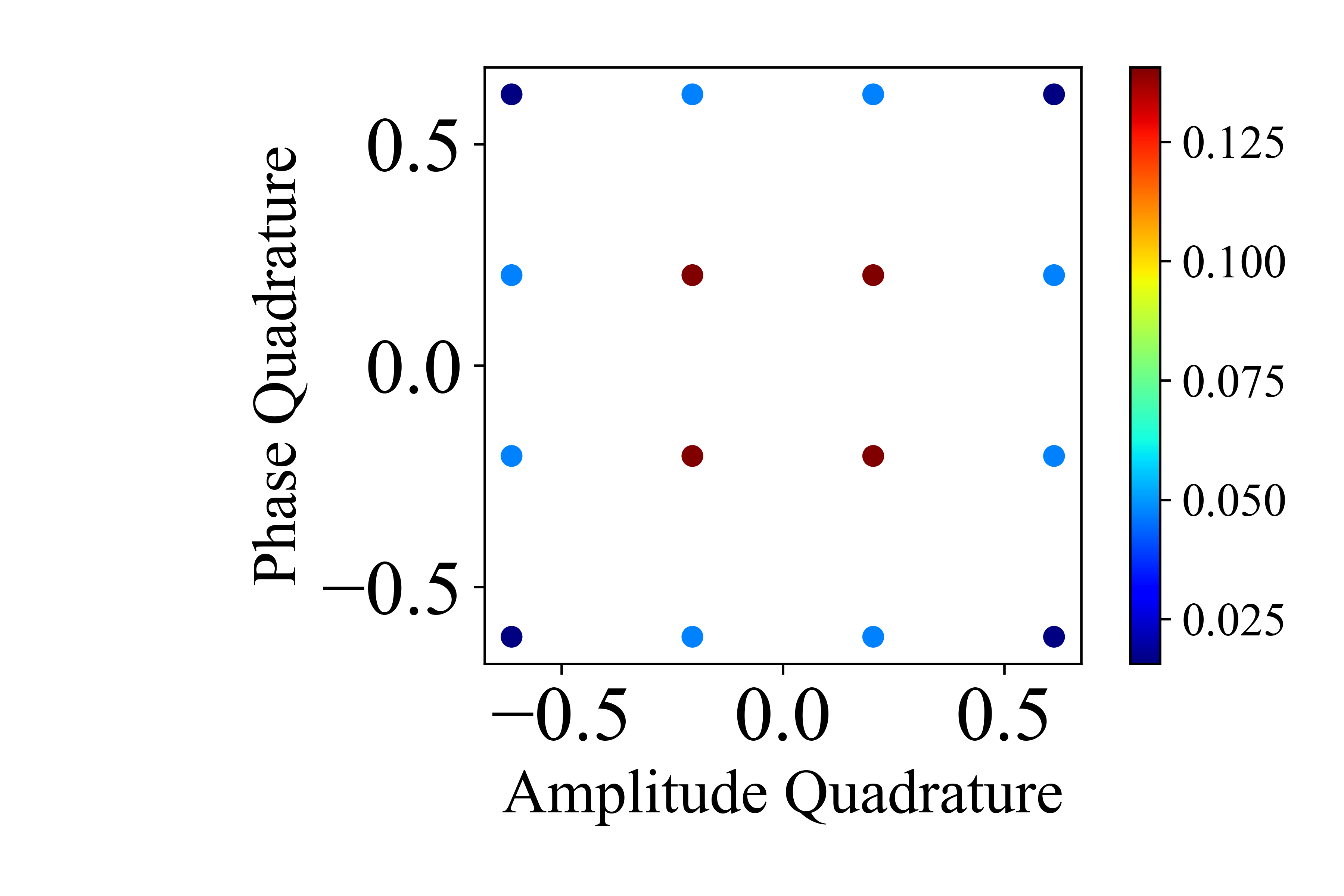}
         \caption{16-QAM}
         \label{fig:16-QAM}
     \end{subfigure}
     \begin{subfigure}[b]{0.43\textwidth}
         \centering
         \includegraphics[width=\textwidth]{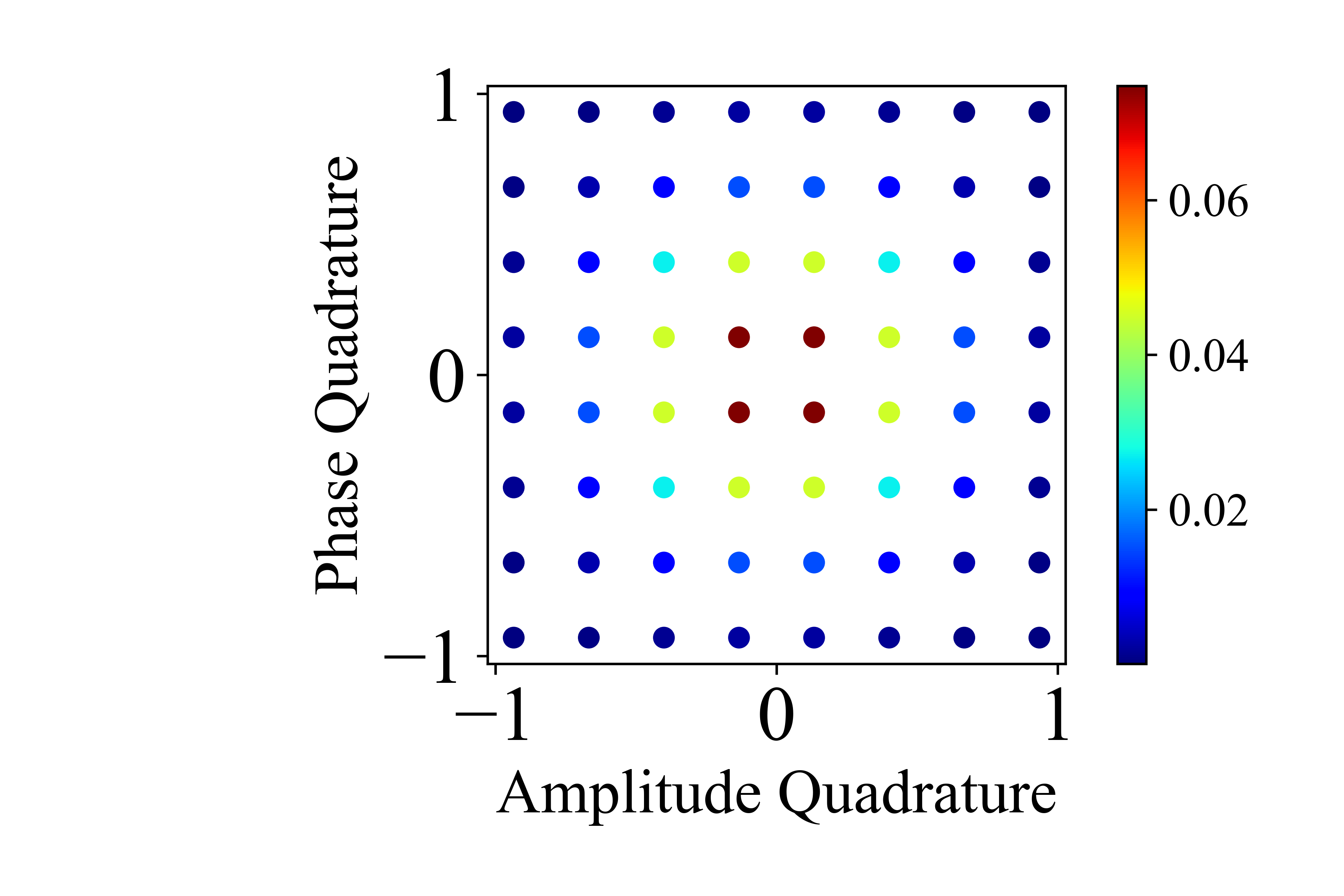}
         \caption{64-QAM}
         \label{fig:64-QAM}
     \end{subfigure}
     \begin{subfigure}[b]{0.43\textwidth}
         \centering
         \includegraphics[width=\textwidth]{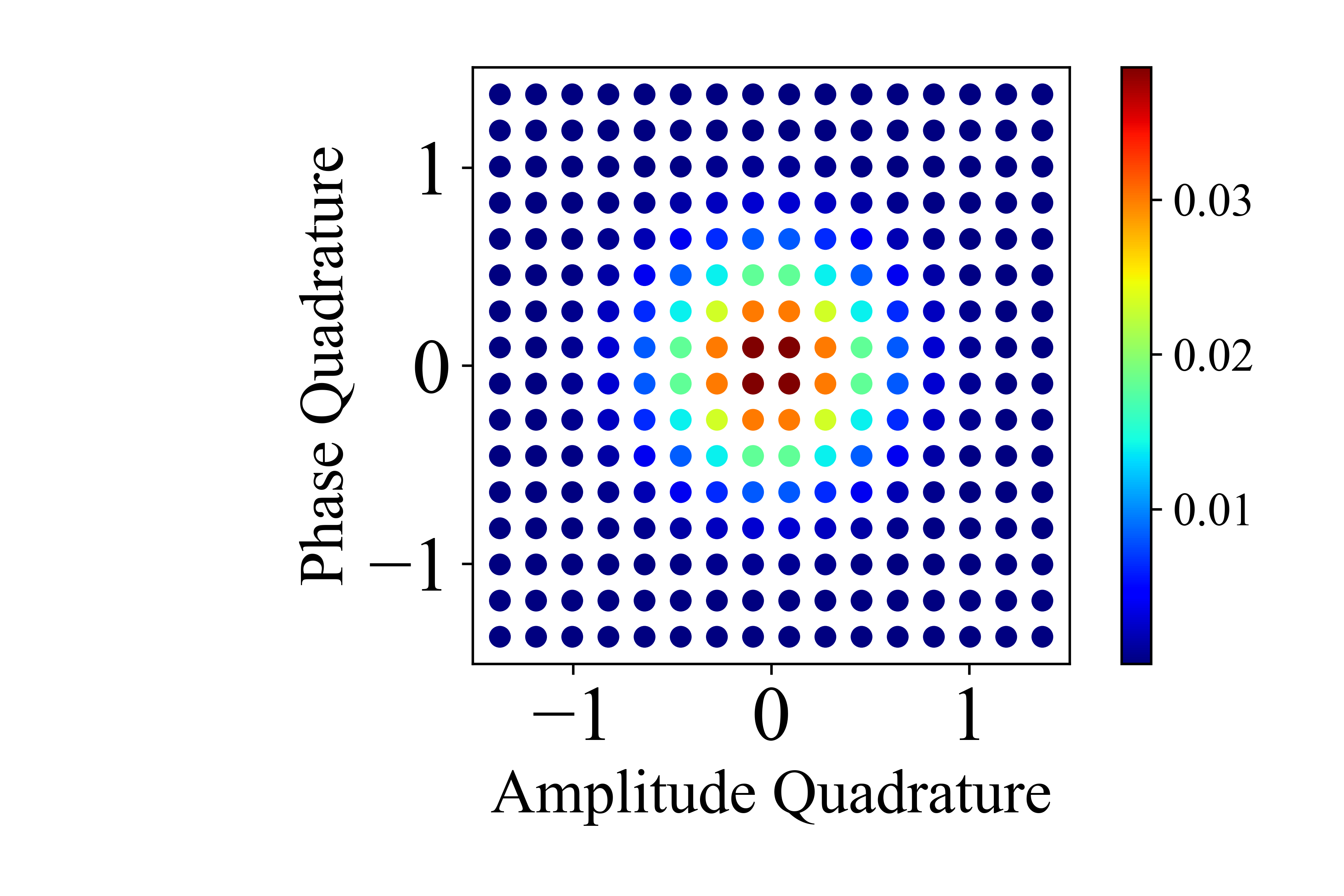}
         \caption{256-QAM}
         \label{fig:256-QAM}
     \end{subfigure}
     \begin{subfigure}[b]{0.43\textwidth}
         \centering
         \includegraphics[width=\textwidth]{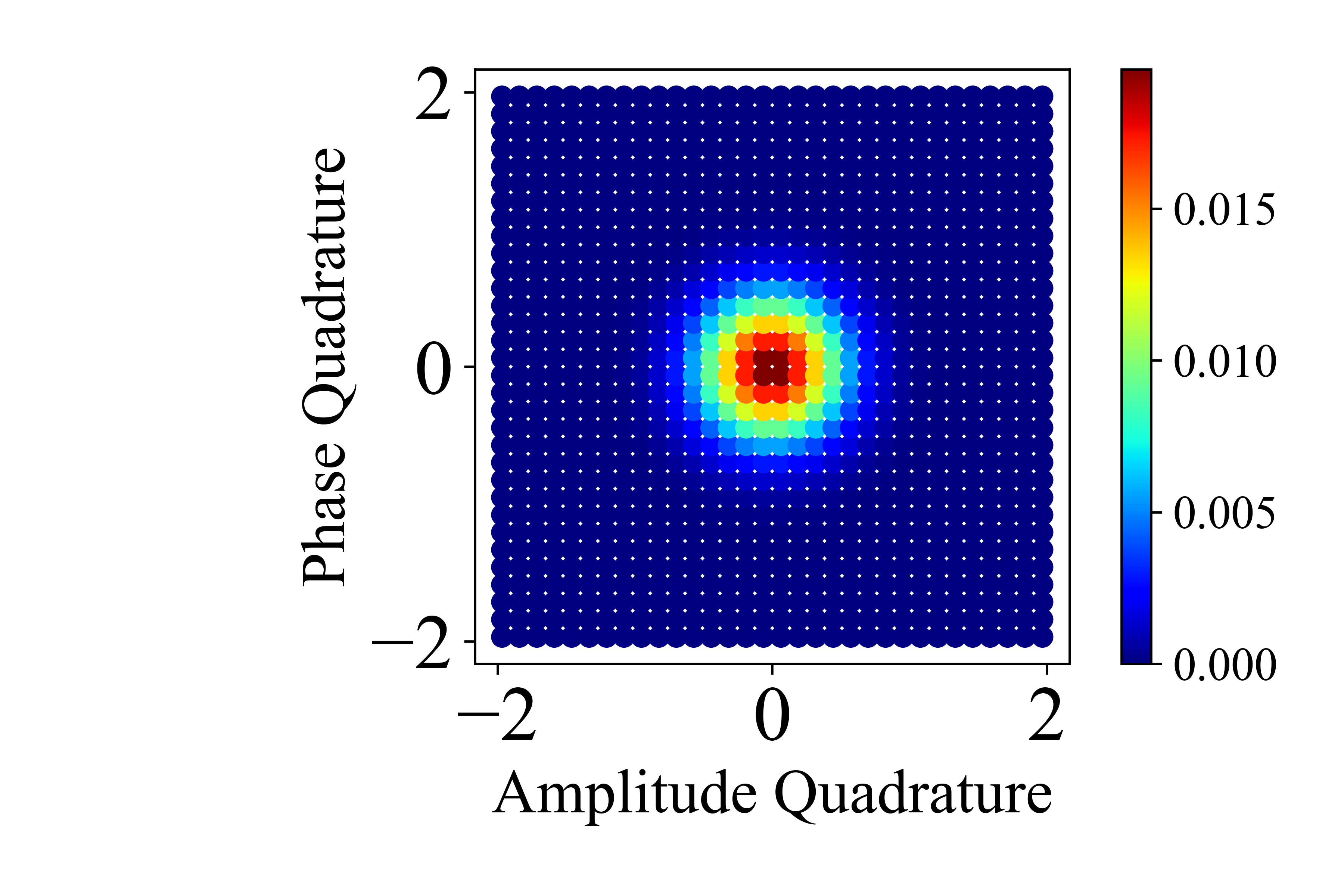}
         \caption{1024-QAM}
         \label{fig:1024-QAM}
     \end{subfigure}
        \caption{Phase space representation of 16,64,256,1024-QAM protocols, $\alpha = 0.5$. Colour bars show the probability of the coherent state for the particular protocol.}
        \label{fig:M-QAM}
\end{figure}

A modulated coherent state takes the form
\begin{equation}
    \label{alpha_QAM}
    \begin{aligned}
        \ket{\alpha_{k, l}} = \frac{\alpha \sqrt{2}}{\sqrt{m - 1}}\left(k - \frac{m - 1}{2}\right) + i\frac{\alpha \sqrt{2}}{\sqrt{m - 1}}\left(l - \frac{m - 1}{2}\right),
    \end{aligned}
\end{equation}
where $M = m^2$ and the coherent states are equidistantly spaced between $-\sqrt{m - 1}$ and $\sqrt{m - 1}$ in the phase and amplitude quadratures. Here,  $k,l \in \{0,1,..,(m-1)\}$. The associated probability for each each coherent state, $p_{k, l}$, can follow either a binomial distribution,
\begin{equation}
    \label{probability_Binomial}
    \begin{aligned}
        p_{k, l} = \frac{1}{2^{2(m - 1)}}\binom{m - 1}{k}\binom{m - 1}{l},
    \end{aligned}
\end{equation}
or a discrete Gaussian distribution,
\begin{equation}
    \label{probability_discreteGaussian}
    \begin{aligned}
        p_{k, l} = \exp(-v(x^2 + p^2)),
    \end{aligned}
\end{equation}
where $x = \frac{\alpha \sqrt{2}}{\sqrt{m - 1}}\left(k - \frac{m - 1}{2}\right)$ and $p = \frac{\alpha \sqrt{2}}{\sqrt{m - 1}}\left(l - \frac{m - 1}{2}\right)$.
\newline
\newline
$M$-QAM on the phase space is depicted as shown in Figure \ref{fig:M-QAM}.
\newline

\subsection{$M$-APSK}
$M$-APSK is a distribution of $M$ modulated coherent states on the phase space where the coherent states are placed in concentric rings. Each ring has a particular number of coherent states \cite{Almeida2021, Almeida2022}. A modulated coherent state takes the form

\begin{equation}
    \ket{\alpha} = \beta_p \alpha \exp(\frac{i2\pi k}{M_p}),
\end{equation}
where $\beta_p = \frac{1}{R}, \frac{2}{R}, ... , 1$ ($R$ is the number of rings), $M_p$ is the number of coherent states in a ring (4, 12, 16, 32, 64, 128, 256 from 1st to 7th ring), and $p = 1, 2, ..., R$. 

A coherent state in a particular ring has equal probability $\left(\frac{1}{M_p}\right)$ and each ring has the same probability $\left(\frac{1}{R}\right)$. Therefore, each coherent state has a probability of $\frac{1}{RM_p}$ in the case of a discrete uniform distribution. 
Other work presents more information about $M$-APSK with non-uniform distributions \cite{Almeida2021, Almeida2022}.

\begin{figure}[htp]
     \centering
     \begin{subfigure}[b]{0.43\textwidth}
         \centering
         \includegraphics[width=\textwidth]{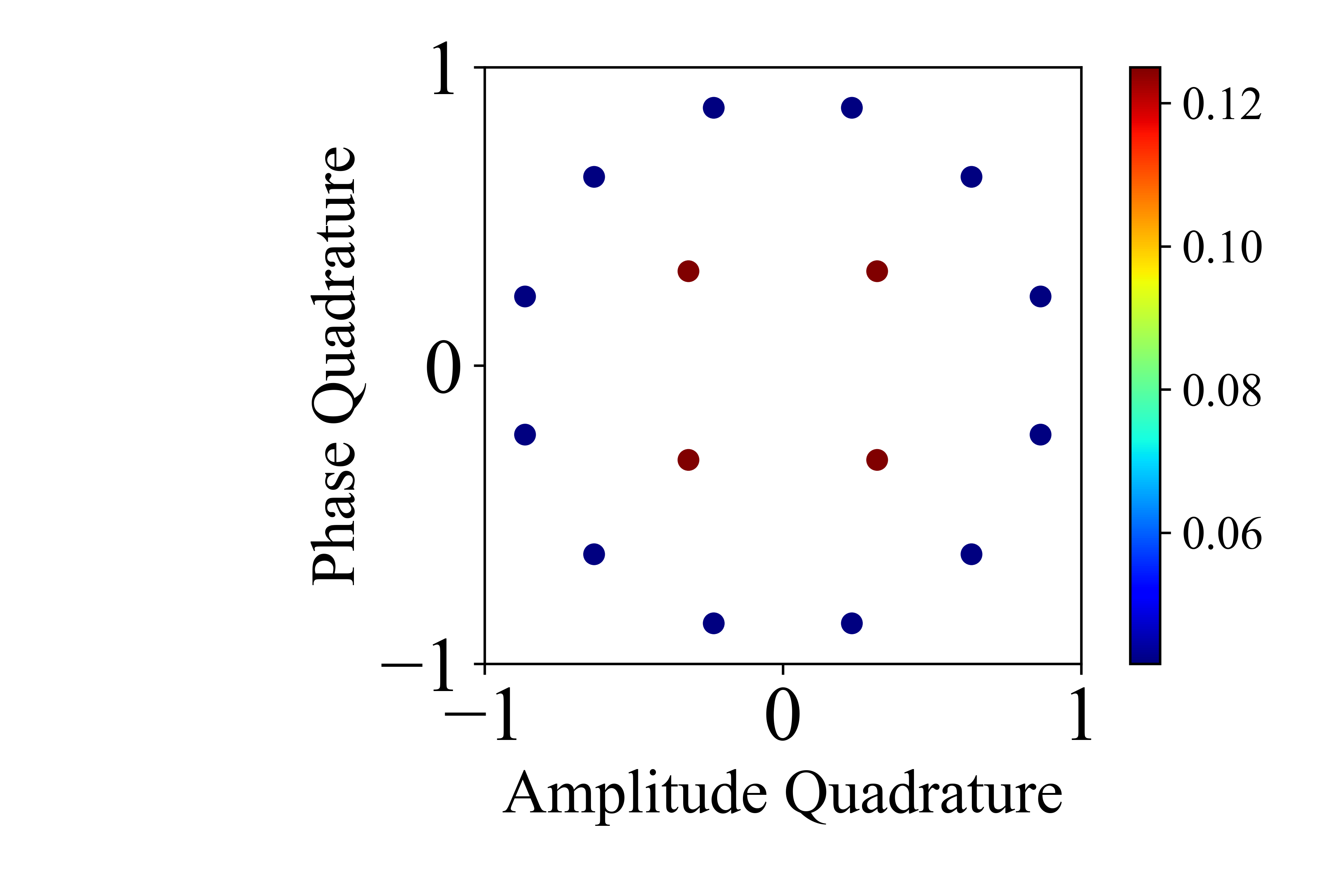}
         \caption{16-APSK}
         \label{fig:16-APSK}
     \end{subfigure}
     \begin{subfigure}[b]{0.43\textwidth}
         \centering
         \includegraphics[width=\textwidth]{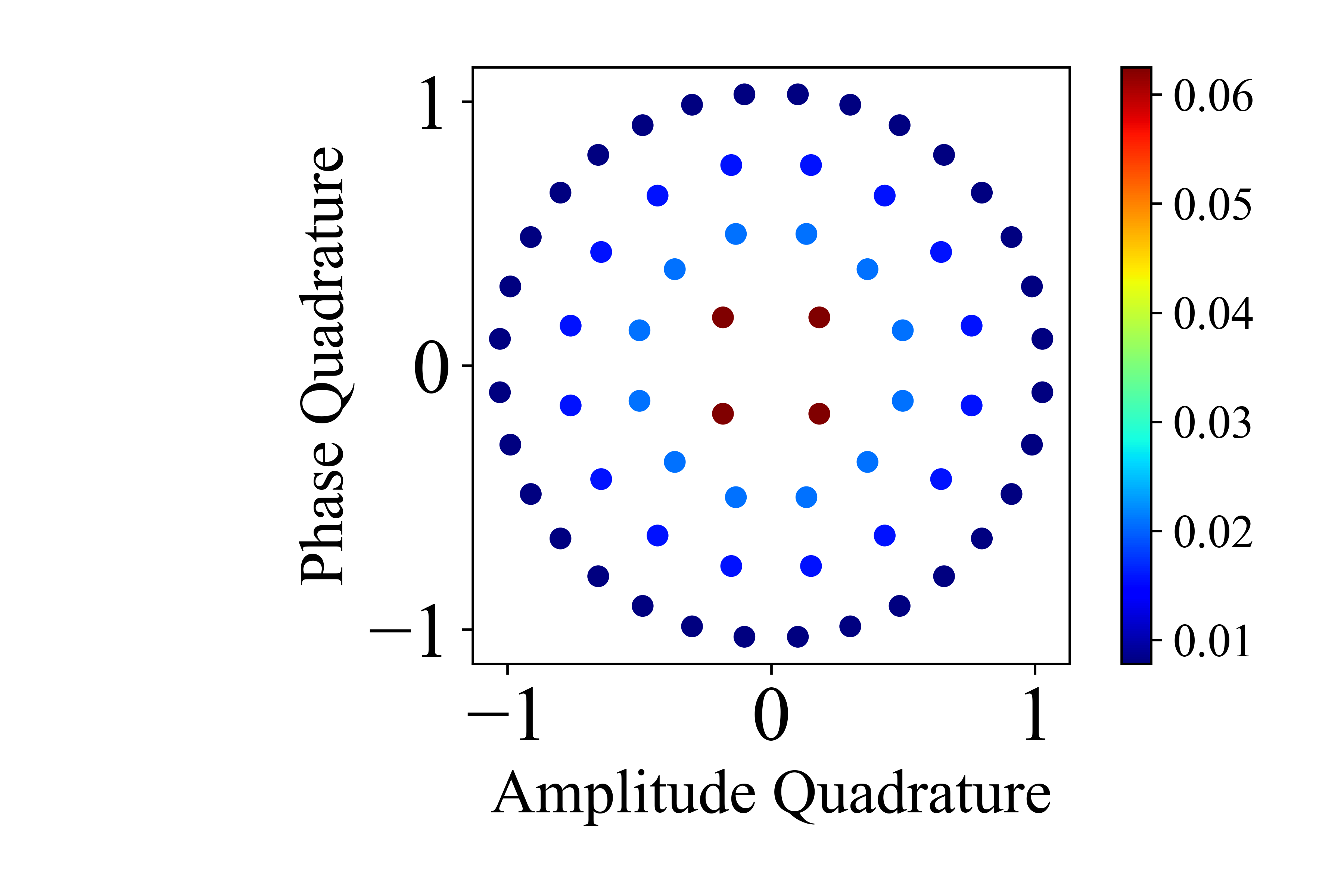}
         \caption{64-APSK}
         \label{fig:64-APSK}
     \end{subfigure}
     \begin{subfigure}[b]{0.43\textwidth}
         \centering
         \includegraphics[width=\textwidth]{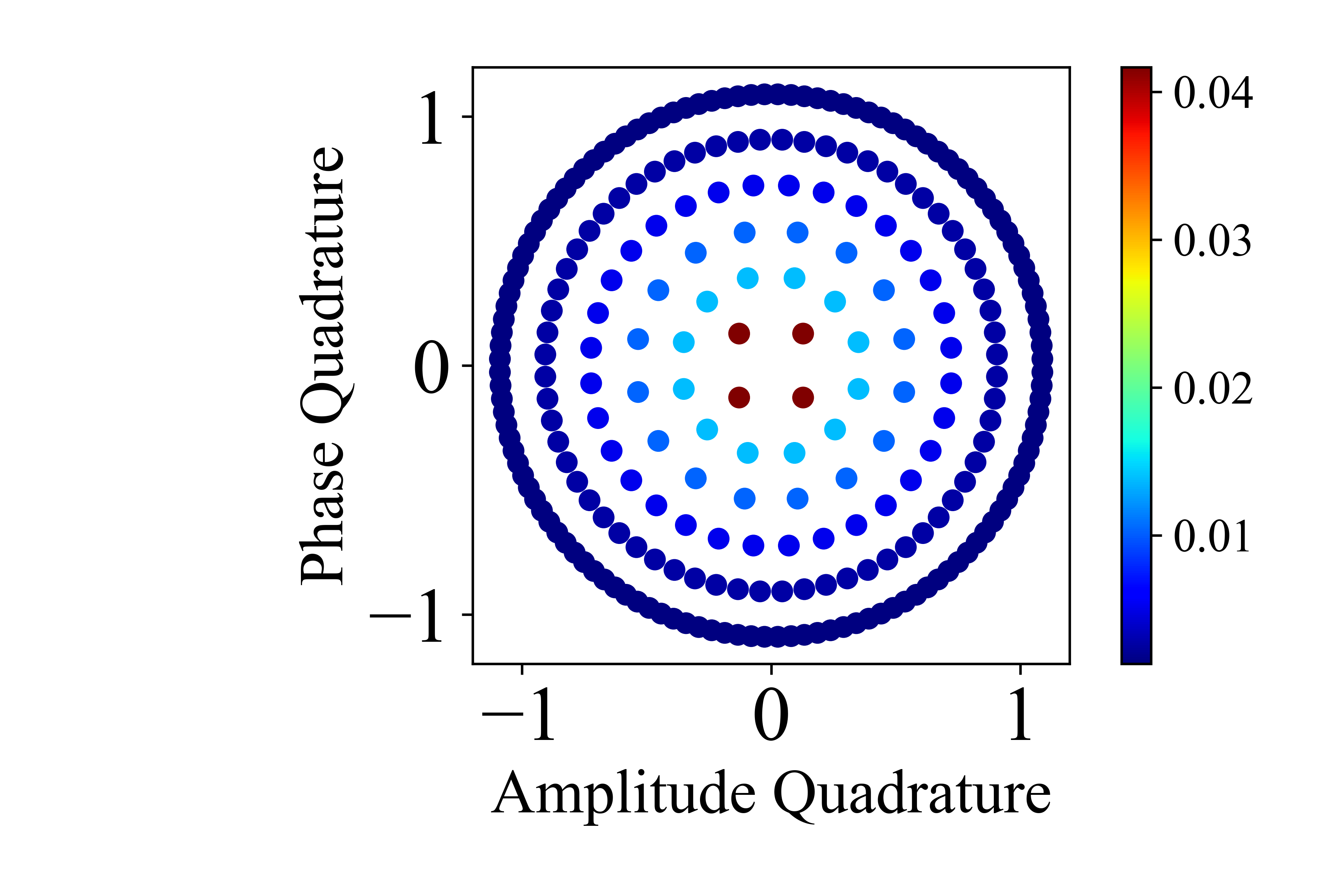}
         \caption{256-APSK}
         \label{fig:256-APSK}
     \end{subfigure}
     \begin{subfigure}[b]{0.43\textwidth}
         \centering
         \includegraphics[width=\textwidth]{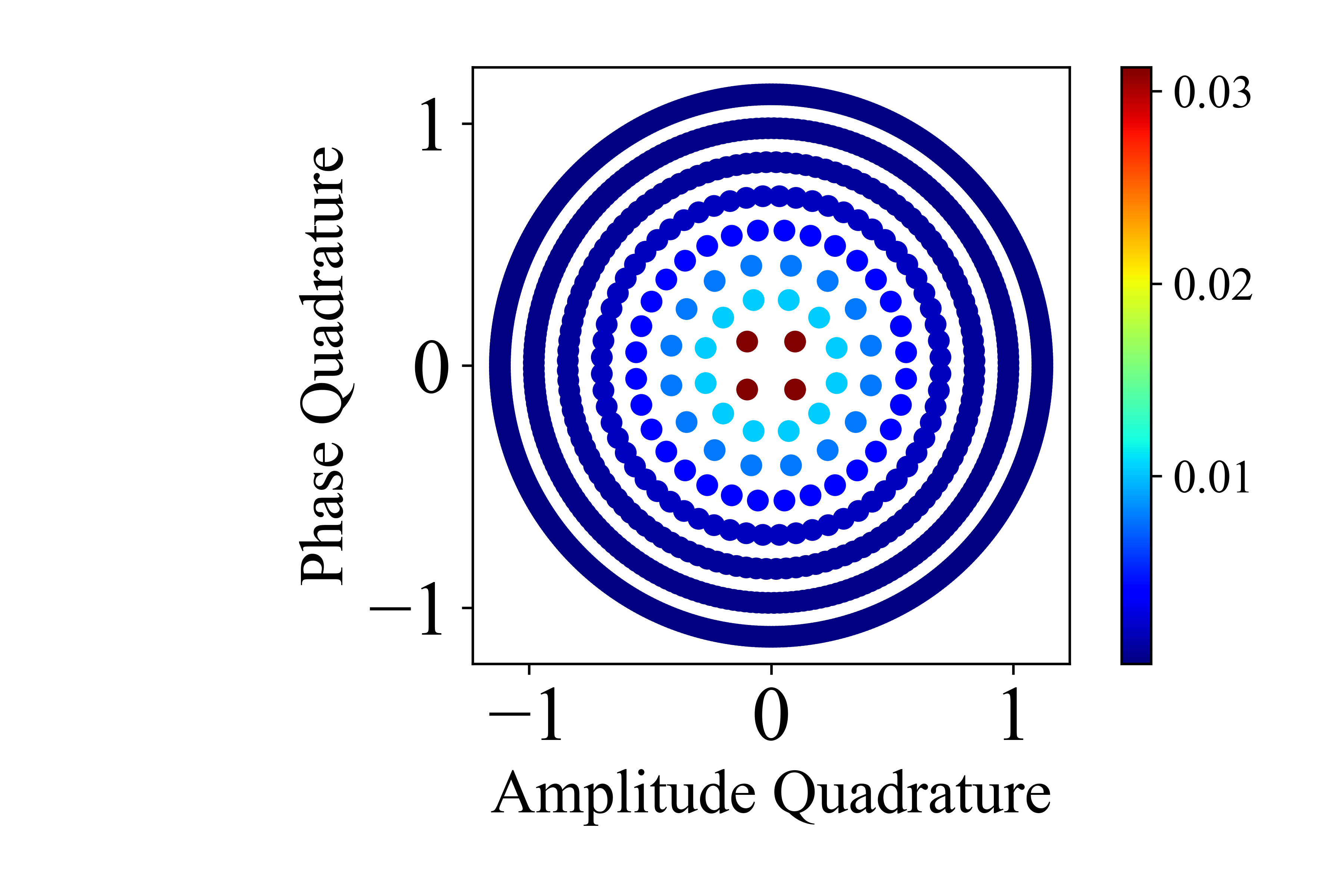}
         \caption{1024-APSK}
         \label{fig:1024-APSK}
     \end{subfigure}
        \caption{Phase space representation of 16,64,256,1024-APSK protocols, $\alpha = 0.707$. Colour bars show the probability of the coherent state for the particular protocol.}
        \label{fig:M-APSK}
\end{figure}

$M$-APSK on the phase space is depicted in Figure \ref{fig:M-APSK}.

\newpage

\section{Method}

The tool for comparing different protocols requires the identification of the boundary between positive and negative SKRs: the minimum positive SKR boundary. This boundary exists in the three-dimensional transmittance--excess-noise--alpha ($T$--$\xi$--$\alpha$) space.  

\begin{figure}[htp!]
    \centering
    \includegraphics[width=0.6\textwidth,keepaspectratio]{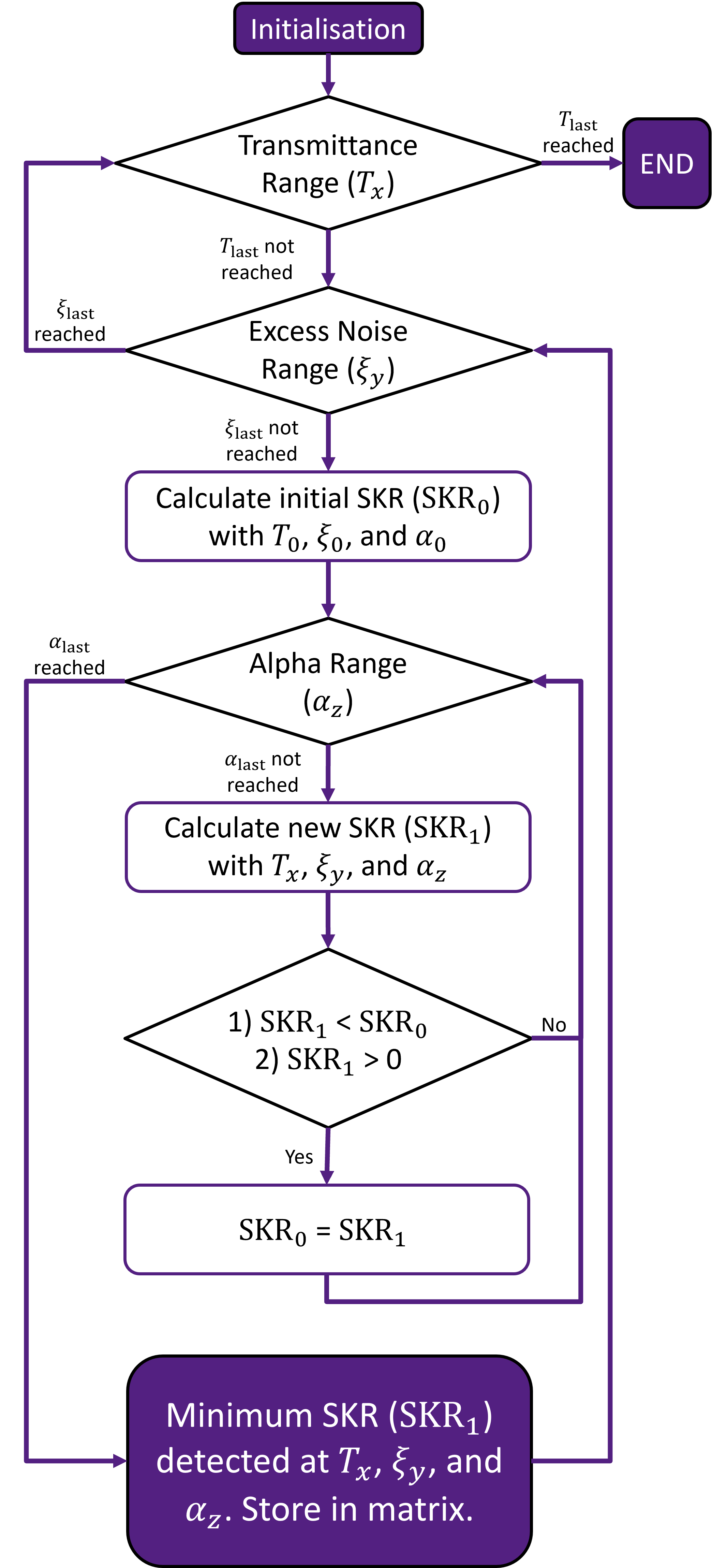}
    \caption{Flowchart for calculating the minimum positive SKR of a protocol for a given $T$, $\xi$, and $\alpha$. The minimum positive SKR points create the minimum positive SKR boundary from which the level of the boundary can be calculated and used to compare different protocols.}
    \label{fig:FlowChart}
\end{figure}

To identify the boundary of the minimum positive SKR for the given parameters, the process described by the flowchart in Figure \ref{fig:FlowChart} is used. A range is defined for $T$, $\xi$, and $\alpha$ to find the minimum positive SKR.

For specified values of $T$ and $\xi$, an initial SKR is calculated using an $\alpha$ value within the defined $\alpha$ range. For each value within the $\alpha$ range, the minimum positive SKR is found by calculating a new positive SKR and comparing it to the initial SKR, or the previous minimum positive SKR. A new minimum positive SKR is found by satisfying two inequalities:

\begin{equation}
\begin{aligned}
    \mathrm{SKR_1} &< \mathrm{SKR_0}, \\
    \mathrm{SKR_1} &> 0,
\end{aligned}
\end{equation}

where $\mathrm{SKR_1}$ is the new calculated SKR within the $\alpha$ range, and $\mathrm{SKR_0}$ is the previous minimum positive SKR or the initial SKR from the first iteration. Once the final $\alpha$ value within the $\alpha$ range is reached, the current minimum positive SKR for the given $T$, $\xi$, and $\alpha$ is stored in a matrix. The process is complete when a minimum positive SKR is found for each $T$, $\xi$, and $\alpha$. A three-dimensional boundary can then be formed containing points of minimum positive SKRs for the defined $T$, $\xi$, and $\alpha$ range. 
\newline
\newline 

\newpage
For comparing different CVQKD protocols, such as the $M$-PSK, $M$-QAM, and $M$-APSK, the $T$, $\xi$, and $\alpha$ are varied. All other parameters required for calculating the SKR are kept constant. The range of each parameter is displayed in Table \ref{Parameters}.

\begin{table} [htp]
\centering
\caption{CVQKD Parameters}
\label{Parameters}
\scalebox{1}{
\begin{tabular}{@{}|ll|@{}}
    \bottomrule
    Parameter &  Value \\ \toprule \bottomrule
    Transmittance ($T$) & 0~--~1 \\
    Excess noise ($\xi$) & 0.001~--~0.5 \\
    Alpha ($\alpha$) & 0.1~--~0.5 \\
    Reconciliation Efficiency ($\beta$) & 0.95 \\
    Detection & Heterodyne \\
    \toprule
\end{tabular}}
\end{table}


The tool's metric for comparing the capability to produce positive SKRs of different protocol is defined as the level of the minimum positive SKR boundary on a three-dimensional $T$--$\xi$--$\alpha$ graph. The level is defined as the average $\alpha$ value, $\alpha_{\mathrm{ave}}$, of the discretised minimum positive SKR boundary from the matrix when the process, described in Figure 4, is complete. A protocol with a minimum positive SKR boundary with a lower level has a larger capability of producing positive SKRs as it has a larger $\alpha$ range, above the boundary, that can produce positive SKRs. 


Creating a three-dimensional minimum positive SKR boundary allows regions of positive and negative SKRs to be identified in three dimensions. In practical operations, this allows the appropriate adjustments to the coherent states (adjustment of $\alpha$ through phase and amplitude modulations) for a given channel.

\begin{figure}[htp]
    \centering
    \includegraphics[width=0.4625\textwidth,keepaspectratio]{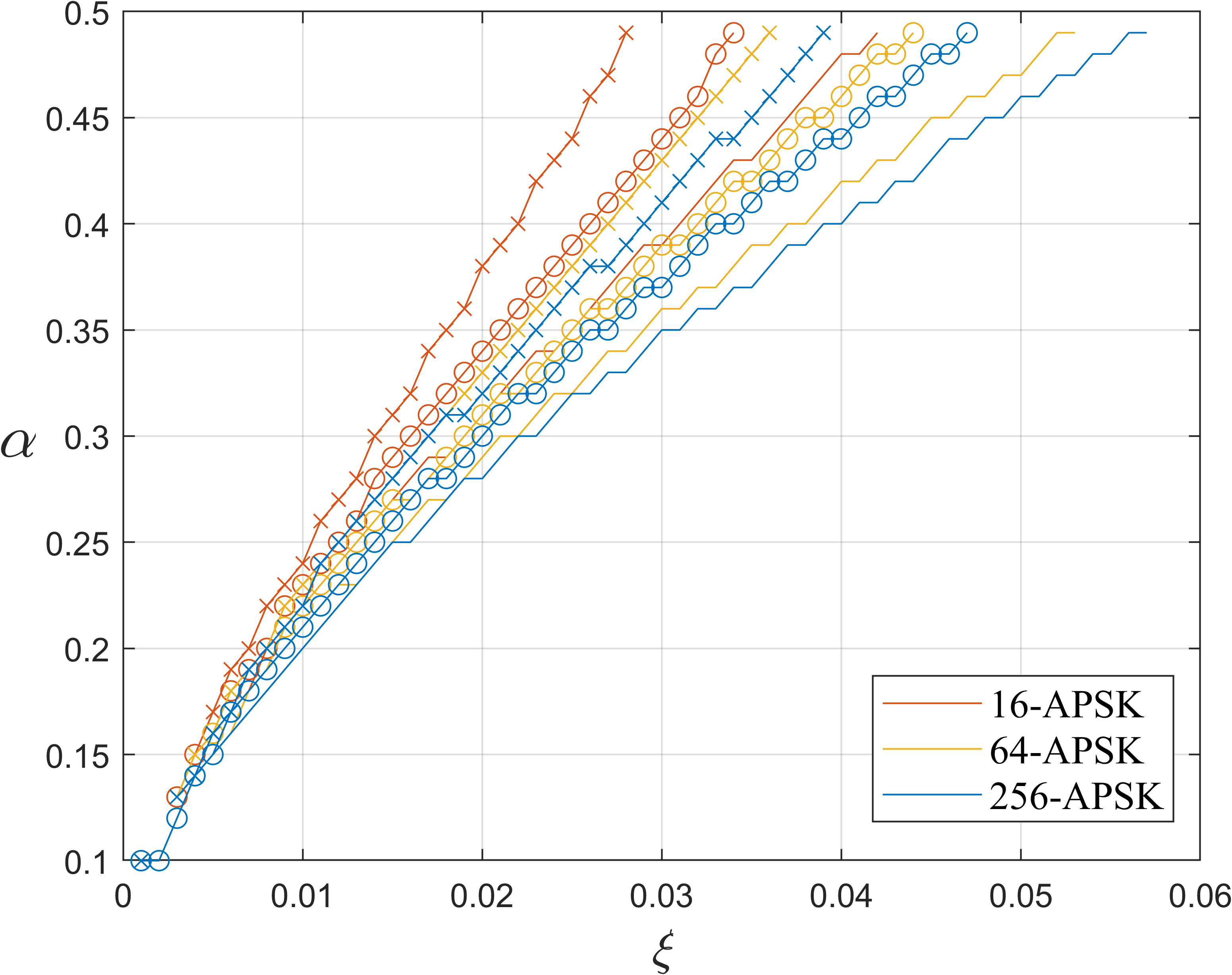}
    \caption{Slices of the minimum positive SKR boundary for the 16 (red), 64 (yellow), and 256-APSK (blue) protocols. $T$ has been set to 0.5 (crossed lines), 0.75 (circled lines), and 1 (solid lines).}
    \label{fig:alphaexcessnoise}
\end{figure}

The usual approach in literature for comparing different protocols is the calculation of the SKR for one parameter ($T$, $\xi$, or $\alpha$ separately) \cite{Becir2010, Zhang2012, Denys2021, Almeida2021, Almeida2022}. However, this restricts comparisons to separate one dimensional metrics, omitting the simultaneous influence of other CVQKD parameters. Figure \ref{fig:alphaexcessnoise} shows the $\alpha$ and $\xi$ required for a minimum positive SKR for the 16, 64, and 256-APSK protocols. $T$ has been set to 0.5, 0.75, and 1. It can be seen that restricting analyses to two dimensions, a parameter (in this case $T$) must be set to constant values to show the simultaneous effects of all parameters ($T$, $\xi$, and $\alpha$) for guaranteeing a positive SKR. However, this leads to restricting the range of values for $T$. Including more values of $T$ would lead to crowding and eventual loss of information. 

Numerical simulations of 16-APSK, 64-APSK, 256-APSK and 16-PSK, 16-APSK, 16-QAM under collective attacks, based on the analytical works of Denys et al. \cite{Denys2021} and Almeida et al. \cite{Almeida2021, Almeida2022}, were performed for intra-protocol and inter-protocol analyses, respectively, to showcase the capabilities of the tool.

\section{Numerical Results}

\begin{figure}[htp]
    \centering
    \includegraphics[width=0.8\textwidth]{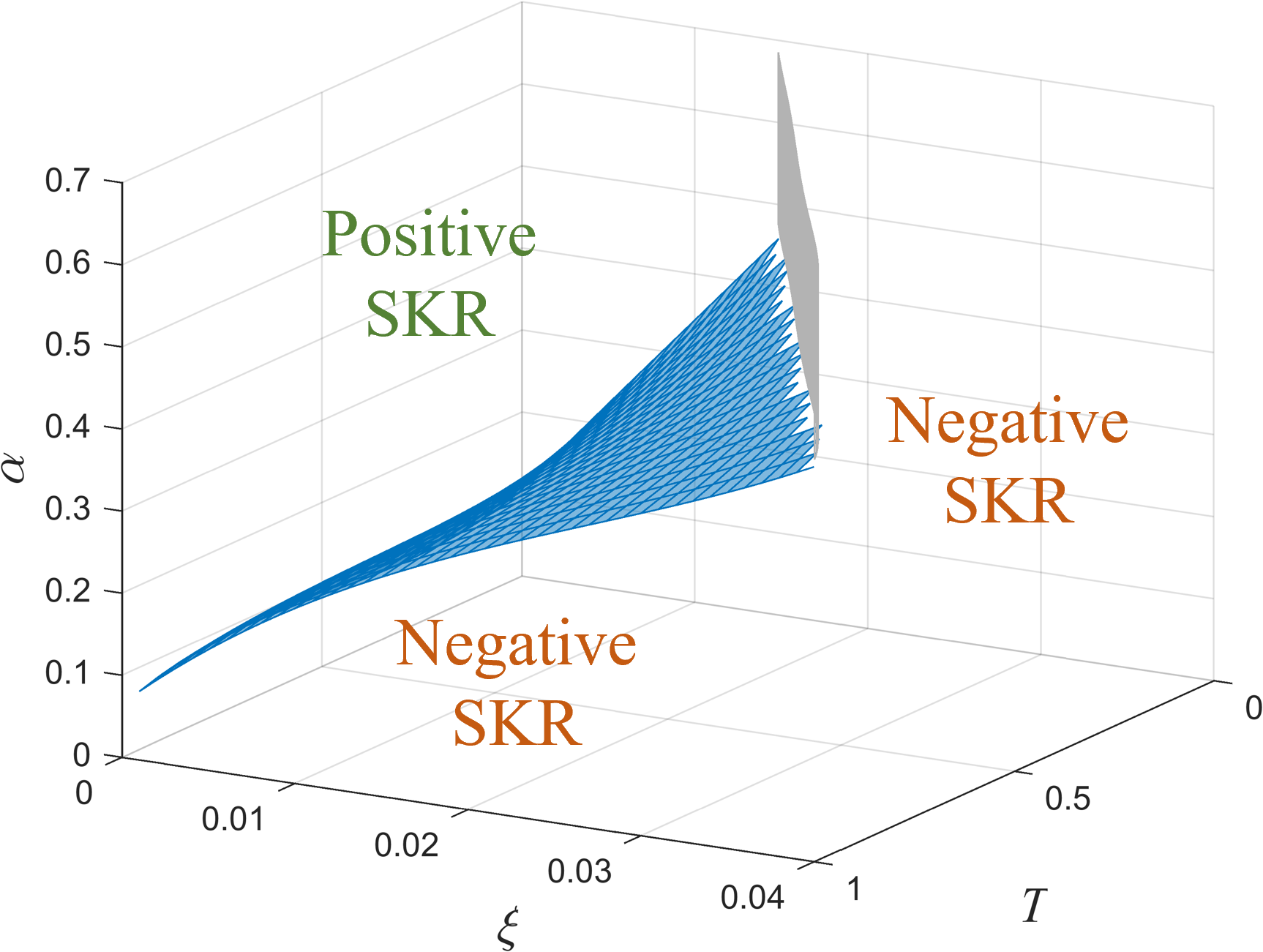}
    \caption{Minimum positive SKR boundary of 16-APSK (blue). The cut-off (grey) is shown to identify regions of positive SKRs and negative SKRs in three dimensions.}
    \label{fig:16QAM_Cutoff_RegionsLabelled}
\end{figure}

A discretised mesh with a surface fit representing the minimum positive SKR boundary is shown in Figure \ref{fig:16QAM_Cutoff_RegionsLabelled} for 16-APSK. Note that the three-dimensional graph can be freely rotated on the three axes. However, snapshots have been taken in this section to best display the profile of the minimum positive SKR boundary.

Figure \ref{fig:16QAM_Cutoff_RegionsLabelled} shows the approximate cut-off beyond which the protocol cannot produce a positive SKR on the $T$--$\xi$ plane. This cut-off changes depending on the protocol but is correctly placed relative to the minimum positive SKR boundary. The cut-off is for the defined $\alpha$ range (Table \ref{Parameters}). If the $\alpha$ range were increased above 0.5, the minimum positive SKR boundary would be extended and a new cut-off placed relatively to it. 

The regions of positive and negative SKR have also been identified. A larger $\alpha$ value produces a more positive SKR \cite{Zhang2012}. It should be noted, however, that another boundary exists past a certain $\alpha$ value. This would close the surface, creating a volume within which positive SKRs are possible. In contrast, negative SKRs exist outside the volume. The volume shape and size would depend upon the protocol.

The surface fit for the discretised mesh is a $\mathrm{3}^{\mathrm{rd}}$ order polynomial that can be expressed as:

\begin{equation}
\label{SurfaceFit}
\begin{aligned}
    \alpha = c_1 + c_2T + c_3\xi + c_4T^2 + c_5T\xi + c_6\xi^2 + c_7T^3 + c_8T^2\xi + c_9T\xi^2 + c_{10}\xi^3;
\end{aligned}
\end{equation}
where $T$ is the transmittance and $\xi$ is the excess noise. The coefficients of the surface fit ($c_i$) are shown in Table \ref{Intra} for intra-protocol comparison and Table \ref{Inter} for inter-protocol comparison. The R-square value has been included as a goodness of fit measure.

\subsection{Intra-Protocol Comparison}
\begin{figure}[htp]
    \centering
    \begin{subfigure}[b]{0.49\textwidth}
         \centering
         \includegraphics[width=\textwidth]{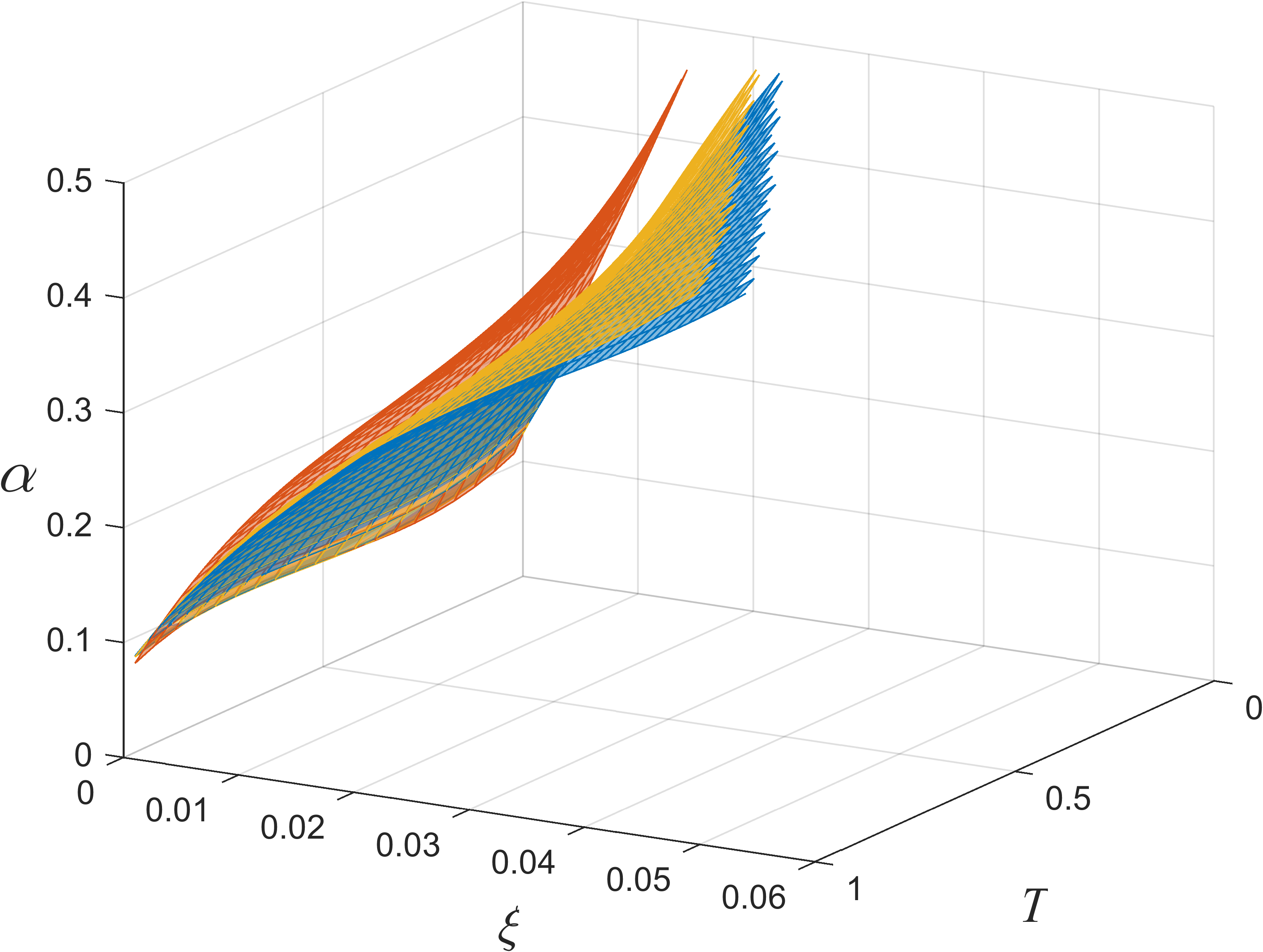}
         \caption{}
         \label{fig:Intra_All}
     \end{subfigure}
     \hfill
     \begin{subfigure}[b]{0.49\textwidth}
         \centering
         \includegraphics[width=\textwidth]{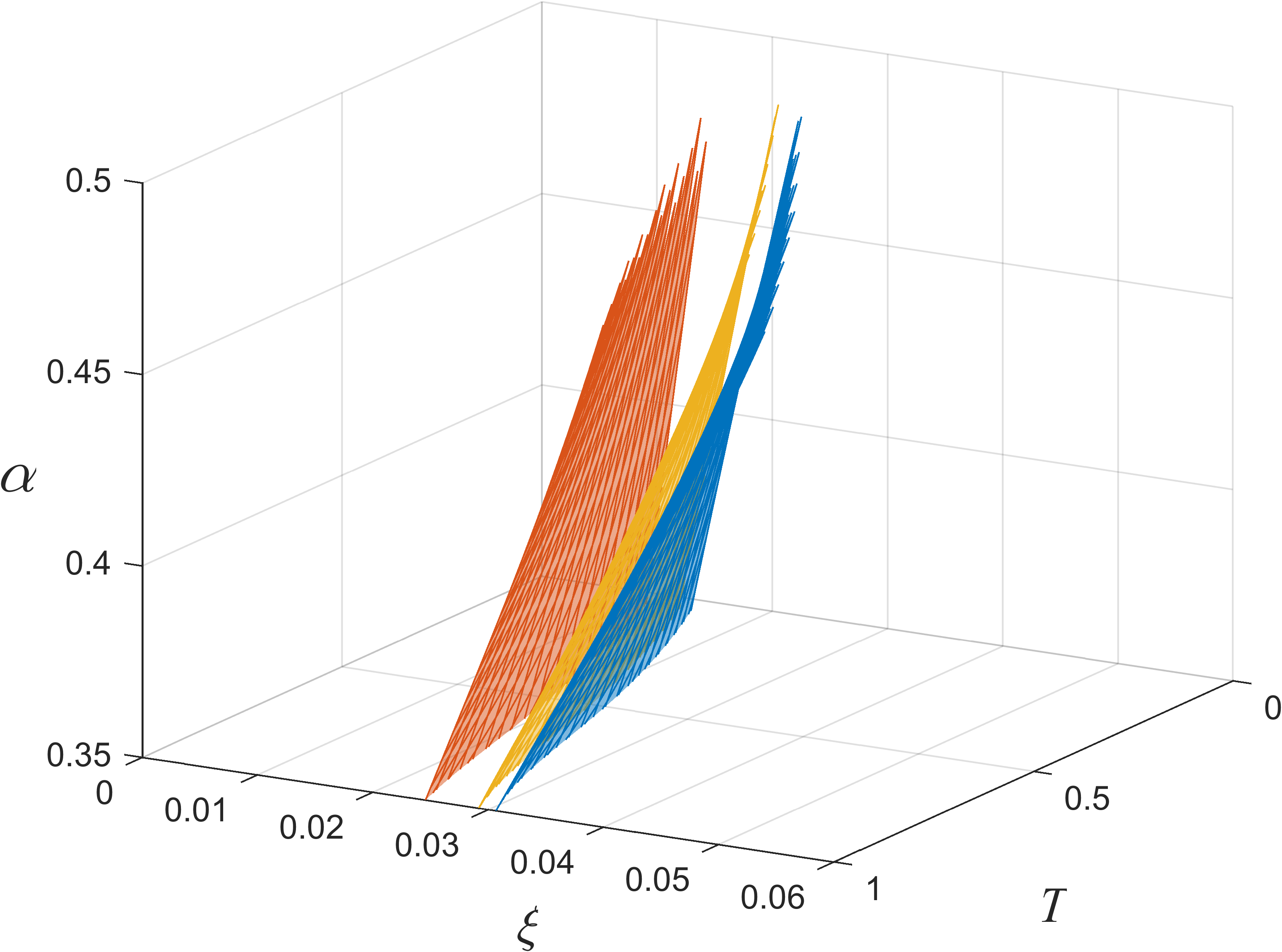}
         \caption{}
         \label{fig:Intra_High}
     \end{subfigure}
     \hfill
    \caption{(a) Minimum positive SKR boundaries for 16-APSK (red), 64-APSK (yellow), and 256-APSK (blue). It can be seen in (b) that the gap between 16-APSK and 64-APSK is larger than the gap between 64-APSK and 256-APSK, showing that the level of the boundary decreases as the number of coherent states ($M$) increases.}
    \label{fig:Intra}
\end{figure}

Using Table \ref{Intra} with Equation \ref{SurfaceFit}, the effect of adding more coherent states in a specific protocol can be compared; in this case 16-APSK, 64-APSK, and 256-APSK. Figure \ref{fig:Intra} shows that increasing the number of coherent states lowers the level of the minimum positive SKR boundary. The larger difference in levels between 16-APSK and 64-APSK compared to 64-APSK and 256-APSK shows that further increases in the number of coherent states for a protocol leads to incremental increases in the capability to produce positive SKRs. However, having a larger number of coherent states has the advantage of having positive SKRs for smaller $T$ and larger $\xi$ values.

For a fair comparison, the calculation of $\alpha_{\mathrm{ave}}$ has been confined to $\xi =$ 0--0.042 SNU and $T =$ 0--1 as this is the region where 16-APSK, the worst performing of the three, produces positive SKRs. The results show that the $\alpha_{\mathrm{ave}}$ of the minimum positive SKR boundaries for 16-APSK, 64-APSK, and 256-APSK are:
\begin{itemize}
    \item 16-APSK: $\alpha_{\mathrm{ave}}$ = 0.304 SNU,
    \item 64-APSK: $\alpha_{\mathrm{ave}}$ = 0.273 SNU, and
    \item 256-APSK: $\alpha_{\mathrm{ave}}$ = 0.265 SNU.
\end{itemize}
It is observed that the $\alpha_{\mathrm{ave}}$ decreases as the number of coherent states increases. In addition, the decrease in $\alpha_{\mathrm{ave}}$ from 64-APSK to 256-APSK is smaller than from 16-APSK to 64-APSK. This is a reflection of the incremental decrease in the level of the minimum positive SKR boundary as the number of coherent states further increases, as shown in Figure \ref{Intra}.

The protocol which has the largest capability to produce positive SKRs is 256-APSK as it has a larger $\alpha$ range, above its minimum positive SKR boundary, to produce positive SKRs by having a lower level minimum positive SKR boundary. It has the smallest $\alpha_{\mathrm{ave}}$ and can produce positive SKRs at larger levels of $\xi$ and lower $T$.

\begin{table}[htp]
\centering
\caption{Numerical Coefficients for 16-APSK, 64-APSK, 256-APSK Surface Fit.}
\label{Intra}
\scalebox{1}{
    \begin{tabular}{|l|r r r|}
        \bottomrule
        Coefficient &  16-APSK & 64-APSK & 256-APSK \\ 
        \toprule 
        \bottomrule
        $c_1$ & 0.1007 & 0.1081 & 0.1114 \\
        $c_2$ & -0.2023 & -0.1963 &	-0.1970 \\
        $c_3$ & 26.09 &	21.39 & 19.85 \\
        $c_4$ & 0.4828 & 0.4083 & 0.4009 \\
        $c_5$ & -24.24 & -13.88 & -11.75 \\
        $c_6$ & -130.1 & -204.8 & -194.2 \\
        $c_7$ & -0.3155 & -0.246 & -0.2396 \\
        $c_8$ & 16.24 &	7.638 & 6.253 \\
        $c_9$ & -250.8 & -67.33 & -49.49 \\
        $c_{10}$ & 4,708 & 2,640 & 2,195 \\
        \hline
        R-Square & 0.9976 & 0.9977 & 0.9978 \\
        \toprule
        
    \end{tabular}}
\end{table}

\newpage
\subsection{Inter-Protocol Comparison}
\begin{figure}[htp]
    \centering
    \begin{subfigure}[b]{0.49\textwidth}
         \centering
         \includegraphics[width=\textwidth]{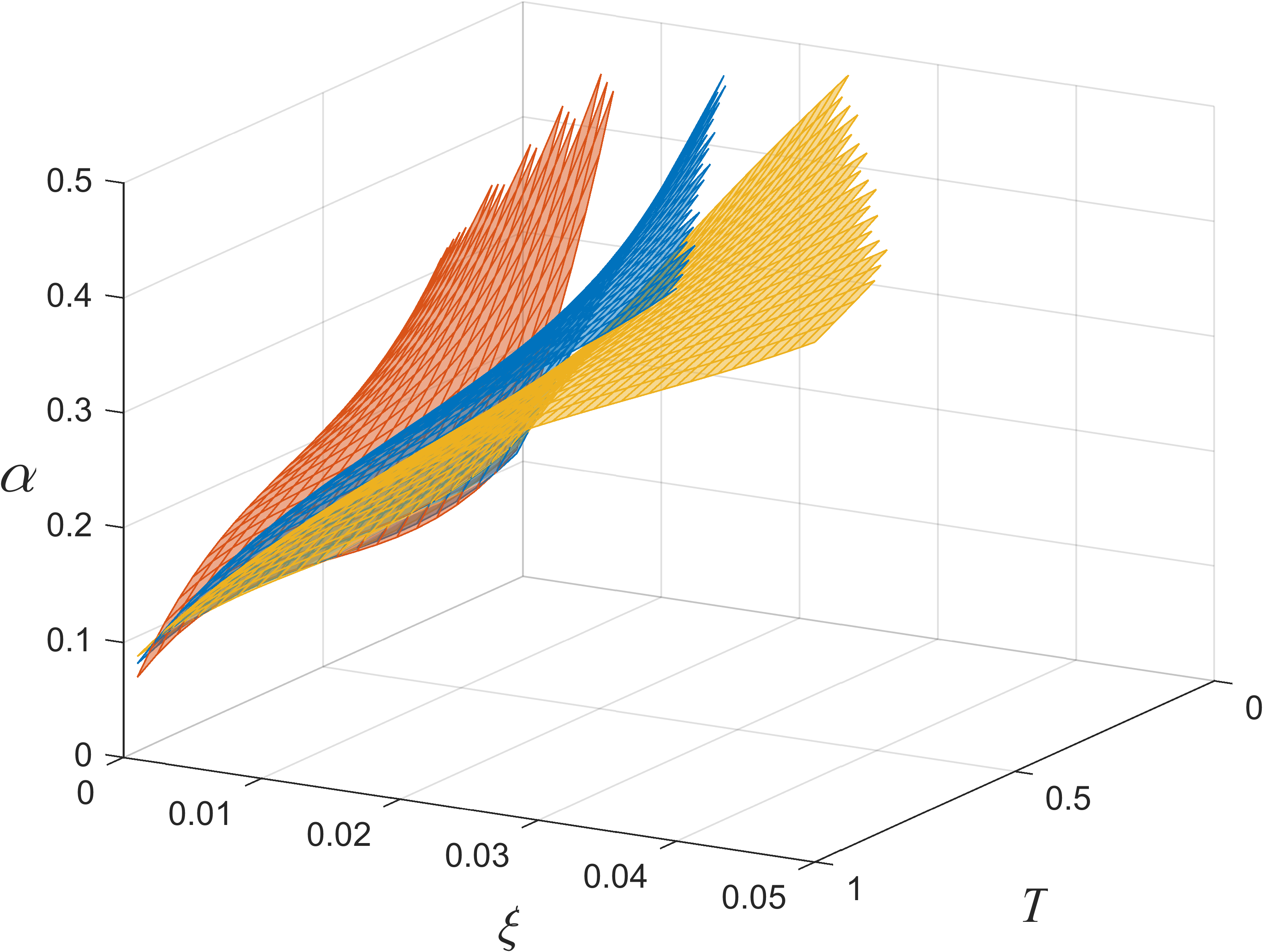}
         \caption{}
         \label{fig:Inter_All}
     \end{subfigure}
     \hfill
     \begin{subfigure}[b]{0.49\textwidth}
         \centering
         \includegraphics[width=\textwidth]{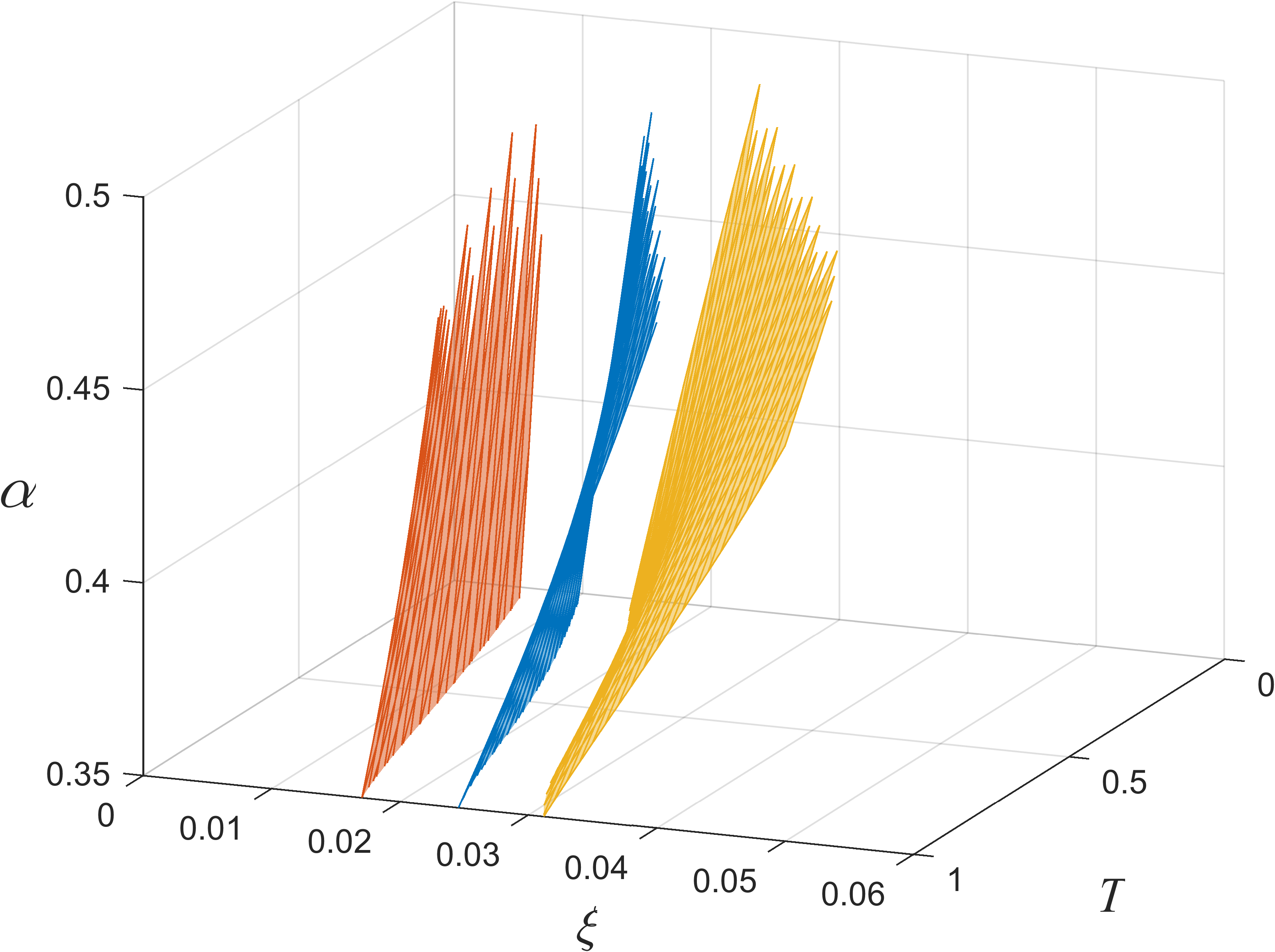}
         \caption{}
         \label{fig:Inter_High}
     \end{subfigure}
     \hfill
    \caption{(a) Minimum positive SKR boundaries for 16-PSK (red), 16-APSK (blue), and 16-QAM (yellow). It can be seen that $M$-QAM is more resilient to unfavourable channel parameters as it has a lower level minimum positive SKR boundary. As seen in (b), $M$-QAM can produce positive SKRs at higher levels of $\xi$.}
    \label{fig:Inter}
\end{figure}

Using Table \ref{Inter} with Equation \ref{SurfaceFit}, the resilience of different protocols can be analysed; in this case, the 16-PSK, 16-APSK, and 16-QAM protocols. The minimum positive SKR boundary for each protocol is shown in Figure \ref{fig:Inter}.

In this case, the calculation of $\alpha_{\mathrm{ave}}$ has been confined to $\xi =$ 0--0.023 SNU and $T =$ 0--1 as this is the region where 16-PSK, the worst performing of the three, produces positive SKRs. The results show that the $\alpha_{\mathrm{ave}}$ of the minimum positive SKR boundaries for 16-PSK, 16-APSK, and 16-QAM are:
\begin{itemize}
    \item 16-PSK: $\alpha_{\mathrm{ave}}$ = 0.256 SNU,
    \item 16-APSK: $\alpha_{\mathrm{ave}}$ = 0.210 SNU, and
    \item 16-QAM: $\alpha_{\mathrm{ave}}$ = 0.193 SNU.
\end{itemize}
The 16-QAM protocol has the largest capability to produce positive SKRs compared to both 16-PSK and 16-APSK by having the smallest $\alpha_{\mathrm{ave}}$. In addition, the 16-QAM minimum positive SKR boundary extends to smaller $T$ and larger $\xi$ values meaning that it has greater resilience in more adverse channels compared to the other protocols. In comparison to Figure \ref{fig:Intra}, it can be seen in Figure \ref{fig:Inter} that the distances between the boundaries for the different protocols are larger. This is evidence that some protocols perform better than others (in this case, $M$-QAM), and that changing protocols may be better (performance-wise) than increasing the number of coherent states in a protocol for a given $T$ and $\xi$ range.

\begin{table}[htp]
\centering
\caption{Numerical Coefficients for 16-PSK, 16-QAM, 16-APSK Surface Fit.}
\label{Inter}
\scalebox{1}{
    \begin{tabular}{|l|r r r|}
        \bottomrule
        Coefficient &  16-PSK & 16-QAM & 16-APSK \\ \toprule \bottomrule
        $c_1$ &	0.09639 & 0.1118 & 0.1007 \\
        $c_2$ &	-0.2603 & -0.1984 & -0.2023 \\
        $c_3$ &	36.92 &	19.61 &	26.09 \\
        $c_4$ &	0.6855 & 0.4022 & 0.4828 \\
        $c_5$ &	-62.34 & -11.44 & -24.24 \\
        $c_6$ &	1203 & -193.4 &	-130.1 \\
        $c_7$ &	-0.4796 & -0.2398 &	-0.3155 \\
        $c_8$ &	56.30 & 5.976 & 16.24 \\
        $c_9$ &	-2614 &	-43.86 & -250.8 \\
        $c_{10}$ &	38,670 &	2,101 &	4,708 \\
        \hline
        R-Square & 0.9822 &	0.9980 & 0.9976 \\
        \toprule
    \end{tabular}}
\end{table}

\subsection{Small Excess Noise Regimes}
\begin{figure}[htp]
     \centering
     \begin{subfigure}[b]{0.45\textwidth}
         \centering
         \includegraphics[width=\textwidth]{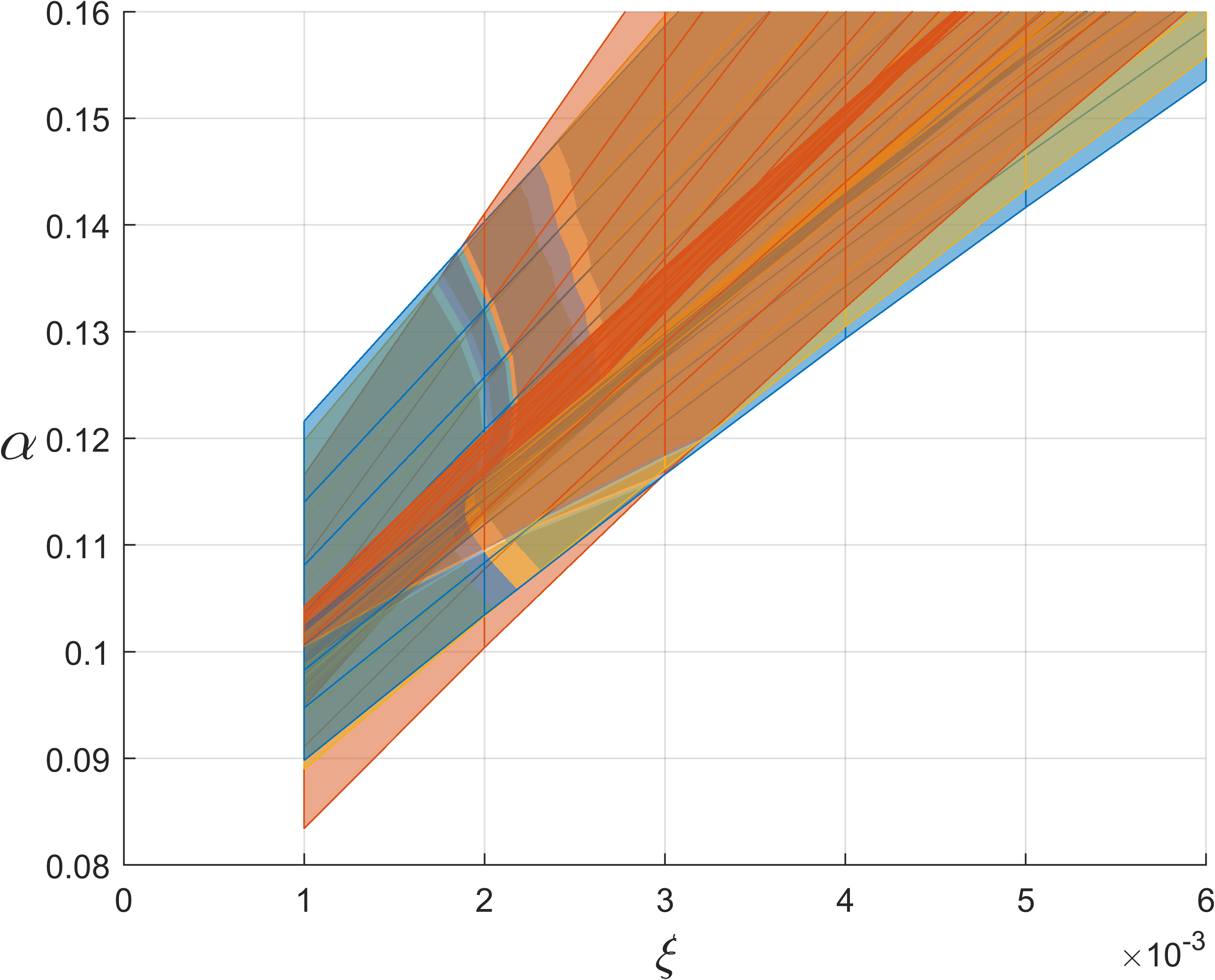}
         \caption{}
         \label{fig:closeupIntra}
     \end{subfigure}
     \hfill
     \begin{subfigure}[b]{0.45\textwidth}
         \centering
         \includegraphics[width=\textwidth]{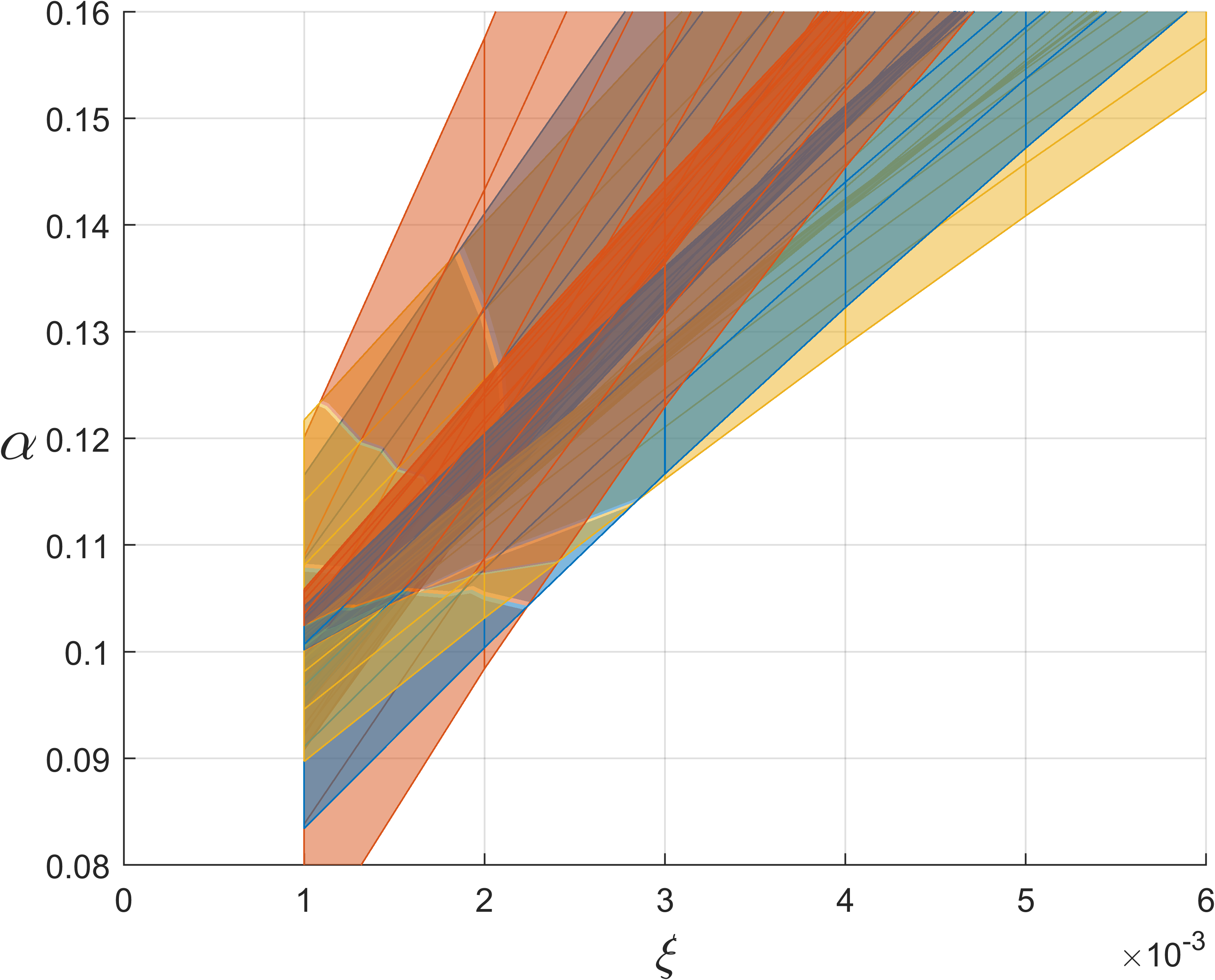}
         \caption{}
         \label{fig:closeupInter}
     \end{subfigure}
     \hfill
    \caption{The minimum positive SKR boundary at small excess noise regimes ($\xi~\sim~10^{-3}$) for (a) intra-protocol [16-APSK (red), 64-APSK (yellow), and 256-APSK (blue)] and (b) inter-protocol [16-PSK (red), 16-APSK (blue), and 16-QAM (yellow)] comparisons. The similar levels of the minimum positive SKR boundaries, shown by the overlaps, indicate similar capabilities for producing positive SKRs at small excess noise regimes.}
    \label{fig:Closeup}
\end{figure}

As shown in Figure \ref{fig:Intra} and Figure \ref{fig:Inter}, the minimum positive SKR boundaries intersect at small excess noise regimes ($\xi~\sim~10^{-3}$) meaning that the capability of $M$-PSK, $M$-QAM, and $M$-APSK to produce positive SKRs are similar for all values of $M$. A closeup of this regime is shown in Figure \ref{fig:Closeup} for all transmittance values ($T = $ 0--1). 

In practical CVQKD implementation, different CVQKD protocols may vary in complexity. The tool can identify regimes where requirements for guaranteeing a positive SKR are relaxed, such as in small excess noise regimes.

\section{Discussion}

The three-dimensional minimum positive SKR boundary can be applied to existing and future security proofs and protocols as a tool to determine the optimum protocol for given channel parameters. Here, only three DM-CVQKD protocols have been compared. In addition, the boundary can be used as a measure of a protocol's resilience in an arbitrary channel. Different channels (ground station to satellite uplink \cite{Hosseinidehaj2016, Pirandola2021, Xu2021}, satellite to ground station downlink \cite{Kish2020, Villasenor2020, Sayat2022}, inter-satellite link \cite{Wang2019, Liu2022}, maritime link \cite{Qu2016, Guo2018, Zhang2019, Mao2020}, terrestrial link \cite{Qu2017, Shen2019} etc.) can be compared for a particular CVQKD protocol by varying the channel parameters.

The tool is also capable of identifying favourable regimes that may relax the requirements for practical CVQKD operation as shown in the case for smaller levels of excess noise. By extension, the tool is also able to identify regimes where the requirements for guaranteeing a positive SKR is more difficult, and may therefore identify the appropriate protocol to produce a positive SKR e.g. larger levels of excess noise. 

The method can be improved by incorporating other CVQKD parameters such as reconciliation efficiency as well as optical and detector efficiencies. Only asymptotic limit SKRs have been analysed. However, finite size limit parameters can also be incorporated such as security parameters, block lengths, and post-processing methods \cite{Johnson2017, Mani2021, Jeong2022}.

In addition, the minimum positive SKR boundary for a given protocol can be extended to a three-dimensional volume containing all positive SKRs. This will include the larger $\alpha$ values which also produces a minimum positive SKR boundary \cite{Zhang2012}. Here, only the smaller $\alpha$ values which produce a minimum positive SKR boundary has been used. 

The shape and size of the volume will have a better indication of a protocol's capability to produce positive SKRs. This is because a lower level (smaller $\alpha_{\mathrm{ave}}$) minimum positive SKR boundary for the smaller $\alpha$ values corresponds to a higher level (larger $\alpha_{\mathrm{ave}}$) minimum positive SKR boundary for the larger $\alpha$ values (Figure 3 in Ref. \cite{Zhang2012}), corresponding to a large sized volume. Conversely, a higher level minimum positive SKR boundary for the smaller $\alpha$ values will have a lower level minimum SKR boundary for the larger $\alpha$ values, corresponding to a smaller sized volume. 

The tool does not yet show the SKR value. In a volume identifying the regions of positive SKR, it would be beneficial to incorporate actual SKR values within the volume. For an arbitrary protocol which could produce SKRs non-linearly with varying channel parameters, the combination of the size and shape of the volume along with actual SKR values may be a better measure for the protocol's resilience and ability to produce positive SKRs.

\section{Conclusions}
A tool for comparing different CVQKD protocols has been presented which considers the simultaneous effects of different CVQKD parameters. Here, the transmittance, excess noise, and alpha have been considered for three different DM-CVQKD protocols. The level of the minimum positive SKR boundary ($\alpha_{\mathrm{ave}}$) has been used as a metric to determine the capability of a protocol to produce positive SKRs. The effects of increasing the number of coherent states for a particular DM-CVQKD protocol shows an increase in the capability to produce positive SKRs. This is shown by a decrease in the level of the minimum positive SKR boundary and a decrease in $\alpha_{\mathrm{ave}}$ as the number of coherent states increase. $M$-QAM outperforms both $M$-PSK and $M$-APSK by having a lower minimum positive SKR boundary level and a smaller $\alpha_{\mathrm{ave}}$. The tool can be adapted by including other CVQKD parameters as well as extending it to a volume containing all positive SKRs where the shape and size of the volume corresponds to a protocol's performance in varying channel parameters and capability to produce positive SKRs.  

\bibliographystyle{IEEEtran}
\addcontentsline{toc}{section}{\refname}\bibliography{References}

\end{document}